\documentclass[usenatbib]{mn2e}
\usepackage{graphicx}
\usepackage{fixltx2e}
\usepackage{times}

\begin{document}
\topmargin-1cm

\newcommand\approxgt{\mbox{$^{>}\hspace{-0.24cm}_{\sim}$}}
\newcommand\approxlt{\mbox{$^{<}\hspace{-0.24cm}_{\sim}$}}
\newcommand\degsq{${\rm\, deg}^2$}

\title[Quasar clustering]{The clustering of intermediate redshift
quasars as measured by the Baryon Oscillation Spectroscopic Survey}
\author[White et al.]{Martin~White$^{1,2}$, Adam~D.~Myers$^{3,4}$,
Nicholas~P.~Ross$^{1}$, David~J.~Schlegel$^{1}$, \newauthor
Joseph~F.~Hennawi$^{4}$, Yue Shen$^{5}$, Ian McGreer$^{6}$,
Michael~A.~Strauss$^{7}$, \newauthor
Adam S. Bolton$^{8}$, Jo Bovy$^{9}$, X. Fan$^{9}$,
Jordi Miralda-Escude$^{10,11}$, N. Palanque-Delabrouille$^{12}$, \newauthor
I. Paris$^{13}$, P. Petitjean$^{13}$, D.P. Schneider$^{14}$, M. Viel$^{15}$,
David H. Weinberg$^{16}$, Ch. Yeche$^{12}$, \newauthor
I. Zehavi$^{17}$, K. Pan$^{18}$, S. Snedden$^{18}$, D. Bizyaev$^{19}$,
H. Brewington$^{18}$, J. Brinkmann$^{18}$, \newauthor
V. Malanushenko$^{18}$, E. Malanushenko$^{18}$, D. Oravetz$^{18}$,
A. Simmons$^{18}$, \newauthor
A. Sheldon$^{18}$, Benjamin A. Weaver$^{19}$ \\
$^{1}$ Physics Division, Lawrence Berkeley National Laboratory,
1 Cyclotron Rd., Berkeley, CA 94720, USA \\
$^{2}$ Departments of Physics and Astronomy, 601 Campbell Hall,
University of California Berkeley, CA 94720, USA \\
$^{3}$ Department of Physics and Astronomy, University of Wyoming,
Laramie, WY 82071, USA \\
$^{4}$ Max-Planck-Institut f\"ur Astronomie, K\"onigstuhl 17, D-69117
  Heidelberg, Germany \\
$^{5}$ Harvard-Smithsonian Center for Astrophysics, 60 Garden Street, MS-51,
Cambridge, MA 02138, USA\\
$^{6}$ Steward Observatory, University of Arizona, 933 N Cherry Ave.,
Tucson, AZ, 85721, USA \\
$^{7}$ Department of Astrophysical Sciences, Princeton University,
Princeton NJ 08544 USA \\
$^{8)}$ Department of Physics and Astronomy, The University of Utah,
  115 S 1400 E, Salt Lake City, UT 84112 \\
$^{9}$ Institute for Advanced Study, Einstein Drive, Princeton, NJ 08540 \\
$^{10}$ Instituci\'o Catalana de Recerca i Estudis Avan\c cats, Barcelona,
Catalonia\\
$^{11}$ Institut de Ci\`encies del Cosmos (IEEC/UB), Barcelona, Catalonia\\
$^{12}$ CEA, Centre de Saclay, IRFU, 91191 Gif-sur-Yvette, France \\
$^{13}$ Universit\'e Paris 6 et CNRS, Institut d’Astrophysique de
Paris, 98bis blvd. Arago, 75014 Paris, France \\
$^{14}$ Dept. Astron. \& Astrophys. and Institute for Gravitation \&
the Cosmos, Pennsylvania State University, University Park, PA 16802 \\
$^{15}$ INAF-Osservatorio Astronomico di Trieste, via G. B. Tiepolo 11,
  I-34131 Trieste, Italy\\
$^{16}$ Department of Astronomy and CCAPP, Ohio State University,
  Columbus, OH, USA \\
$^{17}$ Department of Astronomy, Case Western Reserve University, OH, USA \\
$^{18}$ Apache Point Observatory, P.O. Box 59, Sunspot, NM 88349-0059, USA \\
$^{19}$ Center for Cosmology and Particle Physics, New York University,
  New York, NY 10003 USA
}

\date{\today}
\maketitle

\begin{abstract}
We measure the quasar two-point correlation function over the redshift range
$2.2<z<2.8$ using data from the Baryon Oscillation Spectroscopic Survey.
We use a homogeneous subset of the data consisting of 27,129 quasars with
spectroscopic redshifts---by far the largest such sample used for clustering
measurements at these redshifts to date.
The sample covers 3,600\degsq, corresponding to a comoving volume of
$9.7\,(h^{-1}{\rm Gpc})^3$ assuming a fiducial $\Lambda$CDM cosmology, and
it has a median absolute $i$-band magnitude of $-26$, $k$-corrected to $z=2$.
After accounting for redshift errors we find that the redshift space
correlation function is fit well by a power-law of slope $-2$ and amplitude
$s_0=(9.7\pm 0.5)\,h^{-1}$Mpc over the range $3<s<25\,h^{-1}$Mpc.
The projected correlation function, which integrates out the effects of
peculiar velocities and redshift errors, is fit well by a power-law of slope
$-1$ and $r_0=(8.4\pm 0.6)\,h^{-1}$Mpc over the range $4<R<16\,h^{-1}$Mpc.
There is no evidence for strong luminosity or redshift dependence to the
clustering amplitude, in part because of the limited dynamic range in our
sample.
Our results are consistent with, but more precise than, previous measurements
at similar redshifts.
Our measurement of the quasar clustering amplitude implies a bias factor
of $b\simeq 3.5$ for our quasar sample.
We compare the data to models to constrain the manner in which quasars occupy
dark matter halos at $z\sim 2.4$ and infer that such quasars inhabit halos with
a characteristic mass of $\langle M\rangle\simeq 10^{12}\,h^{-1} M_{\odot}$
with a duty cycle for the quasar activity of 1 per cent.
\end{abstract}

%\keywords{large-scale structure of universe}

\section{Introduction}
\label{sec:introduction}

Quasars are among the most luminous astrophysical objects,
and are believed to be powered by accretion onto supermassive black holes
\citep[e.g.][]{Sal64,Lyn69}.
They have become a key element in our current paradigm of galaxy
evolution -- essentially all spheroidal systems at present harbor massive
black holes \citep{KorRic95}, the masses of which are correlated with many
properties of their host systems.
The emerging picture is that quasar activity and star formation are inextricably
linked \citep[e.g.][]{Nan07,Sil08} in galaxies that contain a massive bulge
(and thus a massive black hole) and a gas reservoir.
The galaxy initially forms in a gas-rich, rotation-dominated system.  Once
the dark matter halo grows to a critical scale some event---most likely a
major merger \citep{Car90,HaiLoe98,CatHaeRee99,KauHae00,Spr05,Hop06} or instability in
a cold-stream fed disk \citep{Cio97,Cio01,DiM12}---triggers a period of rapid,
obscured star formation and the generation of a stellar bulge.
After some time the quasar becomes visible, and soon after the star formation
is quenched on a short timescale, perhaps via radiative or mechanical
feedback from the central engine \citep[e.g.][]{Sha09,Nat12,AleHic12}.

The clustering of quasars as a function of redshift and luminosity provides
useful constraints on our understanding of galaxy evolution.  
The large-scale clustering amplitude increases with the mass of
the dark matter halos hosting the quasars.
Comparison of the abundance of such halos to that of quasars can provide
constraints on the duty cycle and degree of scatter in the observable halo-mass
relation \citep{ColKai89,MarWei01,HaiHui01,WhiMarCoh08,ShaWeiShe10}.
%In particular, the evolution of the host halo mass with redshift is an important constraint
%on models.
However quasars are extremely rare, so very large surveys are necessary to 
suppress the shot-noise from Poisson fluctuations. 
Samples of quasars have only recently included enough objects to study their clustering with some
precision \citep{PorMagNor04,Cro05,PorNor06,Hen06a,Mye06,Mye07a,Mye07b,Ang08,
PWNP,Ros09,Shen++09}.

%Measuring the luminosity dependence of quasar clustering could
%constrain how quasars occupy dark matter halos as a function of 
%their activity (i.e., as a function of Eddington ratio).
%The most luminous quasars in a flux-limited sample are likely to be
%those that harbor massive black holes shining near Eddington.
%But, probing quasar samples to lower luminosity should uncover a
%{\em mix} of more massive, evolved black holes shining far below
%Eddington and less massive black holes shining near
%Eddington \citep[e.g.][]{Lid06,She09,Cao10}.
%The composition of this mix could, in principle, be determined by a precise
%survey of the clustering of quasars at low bolometric luminosity.
%The degree of quasar activity as a function of black hole mass would be of
%particular interest at early times in cosmic history---when feedback from
%quasars is strongly influencing how galaxies evolve.

Naively, measuring the clustering of quasars between redshift 2 and 3 should
be a simple task, as this is where the comoving number density of luminous
quasars seemingly peaks \citep{Wee86,HarSch90,Cro05,Ric06}.
However, selection effects complicate quasar targeting in this range.
The colours of normal (unobscured) quasars around $z\sim 2.7$ resemble far
more abundant stellar populations, particularly metal-poor A and F halo stars
\citep[e.g.,][]{Fan99,Ric01a}.
This issue is enhanced at the faint limits of imaging surveys that achieve
similar depth to the Sloan Digital Sky Survey \citep[SDSS;][]{York00},
where those compact galaxies that are dominated by A and F stellar populations
contaminate selection at the 10--20 per cent level as star-galaxy separation
becomes difficult \citep[e.g.,][]{Ric09,Bov12}.
In addition, faint quasars with $z \sim 2.2$--2.6 can have similar colours to
quasars at $z\sim0.5$, which are contaminated by redder light from their host
galaxy \citep[e.g.,][]{Bud01,Ric01b,Wei04}.
In combination, the cuts that must be made to efficiently select quasars in this redshift range mean that
optical surveys of quasars may miss a significant number of quasars.
A wide-area survey at high targeting density is thus an attractive proposition for the study of quasar
clustering at moderate redshift.
The data we consider in this paper are drawn from just such a survey; 
the Baryon Oscillation Spectroscopic Survey \citep[BOSS;][]{Eis11}.

Physical effects also conspire to make quasars difficult to sample at
$z\sim2.5$.
Obviously, quasars simply appear fainter at greater distances, but in
addition quasars seem to exhibit cosmic downsizing, with the population of less luminous
quasars peaking at lower redshift \citep{Cro09}.
Thus, the most luminous quasars are both more abundant and the most visible
members of the quasar population at $z\sim2.5$.
Quasar clustering measurements at high redshift from the original SDSS
\citep{She07} therefore only sample the most luminous quasars, implying
that deeper spectroscopy than used for the original SDSS quasar survey
\citep{Ric02a} is necessary for sampling quasars across a large dynamic
range in luminosity near $z\sim2.5$.
Indeed, the final BOSS quasar sample should be $\sim10\times$ larger, and
almost 2 magnitudes deeper, than the original SDSS spectroscopic quasar
sample at $2.2 < z < 3.5$ \citep[14,065 objects;][]{Sch10}.

The outline of the paper is as follows.  In \S\ref{sec:data} we describe
the quasar samples that we use---drawn from the SDSS and the BOSS.
The clustering measurements are described in \S\ref{sec:clustering},
including comparisons with earlier work.
The implications of our results for quasars are explored in \S\ref{sec:interp}.
We conclude in \S\ref{sec:conclusions}.
Appendix \ref{sec:zerror} discusses the impact of redshift errors 
on our measurements while
Appendix \ref{sec:halomodel} contains the technical details of the
model fits used in this paper.
Where necessary we shall adopt a $\Lambda$CDM cosmological model with
$\Omega_{\rm mat}=0.274$, $\Omega_\Lambda=0.726$ and $\sigma_8=0.8$
as assumed in \citet{Whi11}, \citet{Aardvark12} and \citet{Rei12}.
Unless the $h$ dependence is explicitly specified or parametrized,
we assume $h=0.7$.
Dark matter halo masses are quoted as $M_{180b}$, i.e.~the mass interior
to a radius within which the mean density is $180\times$ the background density
of the Universe.
Luminosities will be quoted in Watts, and magnitudes in the AB system.

\section{Data}
\label{sec:data}

The Sloan Digital Sky Survey \citep[SDSS;][]{York00} mapped nearly a quarter
of the sky using the dedicated Sloan Foundation 2.5-meter telescope \citep{Gun06}
located at Apache Point Observatory in New Mexico.
A drift-scanning mosaic CCD camera \citep{Gun98} imaged the sky in
five photometric bandpasses \citep{Fuk96,Smi02,Doi10}
to a limiting magnitude of $r\simeq 22.5$.
The imaging data were processed through a series of pipelines that perform
astrometric calibration \citep{Pie03},
photometric reduction \citep{Photo},
and photometric calibration \citep{Pad08}.
We use quasars selected from this 5-band SDSS photometry as described in
detail in \citet{Bov11} and \citet{Ros12}. 

Selecting quasars in the redshift range
$z\simeq 2$--3, where the space density of the brightest quasars peaks
\citep{Ric06,Cro09}, is made difficult by the large populations of
metal-poor A and F stars, faint lower redshift quasars and compact
galaxies which have similar colours to the objects of interest
\citep[e.g.,][]{Fan99,Ric01a}.
In our case the problem is compounded by the fact that we wish to work
close to the detection limits of the SDSS photometry, where errors on
flux measurements cause objects to scatter substantially in colour space,
and that the BOSS key science programs did not include the study of the
clustering of $z\sim 2.5$ quasars.

\begin{figure}
\begin{center}
\resizebox{1.7in}{!}{\includegraphics[bb=115 205 550 599,angle=270]{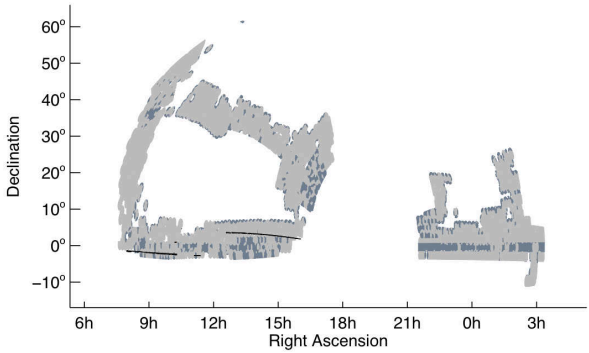}}
\end{center}
\caption{The angular distribution of our quasar sample, in J2000 equatorial
coordinates and Aitoff projection.  We have rotated the reference
line by $90^\circ$ so that the North and South Galactic survey regions
appear contiguous in the left and right parts of the plot,
respectively. Areas which we use in our analysis (light grey), 
have completeness to XDQSO targets of greater than 75 per cent.
Other areas (dark grey) are mainly early survey regions where XDQSO was
not used as the CORE targeting algorithm.  The black areas depict imaging data in
which the $u$-band chip was not operating,
which are discarded from our analysis.}
\label{fig:angular}
\end{figure}

\subsection{Clustering subsamples and the angular mask}
\label{sec:datamask}

The quasar component of BOSS is designed primarily as a Lyman-$\alpha$ Forest
survey, which does not require quasars to be selected in a uniform---or
even a {\it recreatable}---manner across the sky.
For the study of quasar clustering however, uniform selection is key.
To satisfy these competing scientific requirements, the survey thus
adopted a CORE+BONUS strategy, where the CORE objects correspond
to a uniform sample selected by the
{\it extreme deconvolution} (XD)\footnote{XD \citep{Bov09} is a method to
describe the underlying distribution function of a series of points in
parameter space (e.g., quasars in colour space) by modeling that distribution
as a sum of Gaussians convolved with measurement errors.} algorithm. XD is 
applied in BOSS to model the distributions of quasars and stars in flux space, and hence to
separate quasar targets from stellar contaminants \citep[XDQSO;][]{Bov11}.
It is this CORE sample that we analyze in this paper.

Specifically, we take as quasar targets all point sources in SDSS imaging
that have an XDQSO probability above a threshold of $0.424$ to the
magnitude limit of BOSS quasar target selection \citep[$g\leq22.0$ or $r\leq 21.85$;][]{Ros12}.
By quasar targets in this sense, we mean all quasars that would have been
observed in a perfect survey. In reality, not all such targets are observed
because not all fibers can be placed on an object during normal survey
operations. More importantly, the XDQSO algorithm was not adopted as the
final CORE algorithm for BOSS quasar target selection until the second year of
operations \citep{Ros12}. We will work with spectroscopy
take on or before January 1, 2012---just over the first two years of BOSS data.
So, on average, more potential targets with an 
XDQSO probability greater than $0.424$ are unobserved (spectroscopically)
in areas that were covered in the first year of BOSS
(see Fig.~\ref{fig:angular}).

\begin{figure}
\begin{center}
\resizebox{3.4in}{!}{\includegraphics{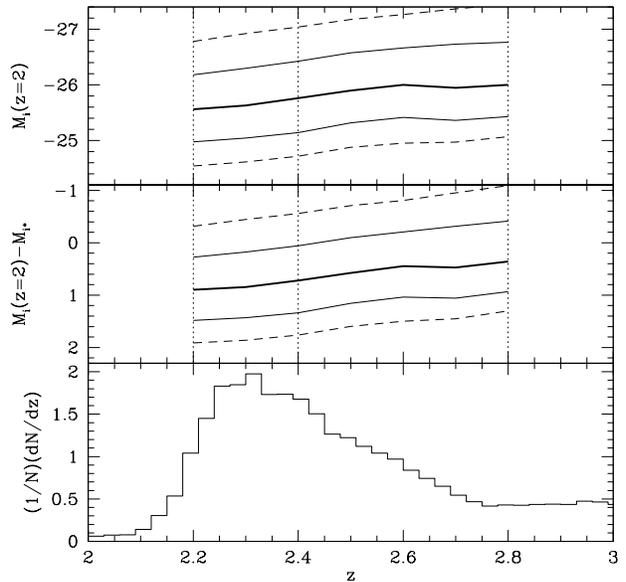}}
\end{center}
\caption{The absolute magnitude distribution and number of quasars vs.~redshift
for our sample.  (Upper) The $10^{\rm th}$, $25^{\rm th}$, $50^{\rm th}$,
$75^{\rm th}$ and $90^{\rm th}$ percentiles of $M_i$ vs.~redshift (see text).
(Middle) The same percentiles now in $M_i-M_{\star,i}$ vs.~redshift.
(Lower) the (normalized) redshift distribution of quasars.
The vertical dotted lines indicate the redshift ranges we consider in our
study.}
\label{fig:magz}
\end{figure}

We use the {\sc Mangle\/} software \citep{Mangle} to track the angular
completeness of the survey (the mask).
The completeness on the sky is determined from the fraction of quasar
targets in a sector for which we obtained a spectrum; a sector is
an area of the sky covered by a unique set of spectroscopic tiles
\citep[see][]{Tiling,Teg04}. For our analyses, we limit the survey to areas
with targeting completeness greater than 75 per cent.
By targeting completeness, we mean the ratio of the number of quasar
targets that received a BOSS fiber to
{\em all} quasar targets. 

%\begin{figure}
%\begin{center}
%\resizebox{3.4in}{!}{\includegraphics{mollweide.eps}}
%\end{center}
%\caption{The angular distribution of our quasar sample, in ecliptic
%coordinates and Mollweide projection.  We have rotated the reference
%line by $90^\circ$ so that the northern and southern regions of the
%survey appear contiguous, in the left and right parts of the plot
%respectively.}
%\label{fig:angular}
%\end{figure}

We do not correct for {\em spectroscopic} incompleteness---i.e.~account
for the fraction of observed targets which produce a spectrum of sufficient
quality to measure a redshift.
Quasars at $z > 2.2$ are identifiable in BOSS even at very low signal-to-noise
ratio because the strong Ly\,$\alpha$\,$\lambda$1215 line always falls within
the BOSS wavelength coverage \citep[3,600--10,000\AA;][]{Eis11}.
In BOSS, almost all unidentifiable objects are likely to be stars, galaxies
or low redshift quasars, not the $z > 2.2$ quasars of interest in this paper.
Correcting for spectroscopic incompleteness would induce a false large scale
clustering signal because the density of stellar contaminants varies over
the sky.

%We should update this and include
%discussion with numbers from truth table exercises we've done.

We apply a veto mask to remove 
survey regions in which a quasar could never be observed---
areas near bright stars and the centerposts of the spectroscopic plates 
\citep[as described in][]{Whi11}. We also remove fields where the 
conditions were not deemed photometric by the SDSS imaging pipeline 
\citep[again see][]{Whi11}. Finally, we remove areas that have
bad $u$ data in the SDSS imaging scans \citep[\S3.3 of][see Fig.~\ref{fig:angular}]{DR2}.
The resulting XDQSO CORE targets were matched to the list of objects for which the
BOSS successfully obtained a spectrum, and throughout this paper
we only consider regions where at least 75 per cent of the XDQSO CORE
targets received a BOSS fiber for spectroscopic observation.

We study BOSS data taken on or before January 1, 2012.
This limits our analysis 
to version 5\_5\_0  of the spectral reduction pipeline
({\tt spAll-v5\_5\_0}) and to 
the areas plotted in Fig.~\ref{fig:angular}.
Note that these are slightly later reductions than made publicly available
with SDSS Data Release 9 (DR9; which uses v5\_4\_45).
Algorithmically v5\_5\_0 is the same as v5\_4\_45, but changes for
calibrations of a newly installed CCD affects data past DR9.

\subsection{Redshift assignation}

The BOSS wavelength coverage is 3,600--10,000\AA\ so for quasars above
$z\sim1$ the [O\,III]\,$\lambda\lambda$4958,5007 complex is shifted out
of the BOSS window.
Systemic redshifts for most BOSS quasars thus rely solely on information from
broad emission lines in the rest-frame ultraviolet.
The next lowest ionization line typically found in quasar spectra
Mg\,II\,$\lambda$2798, which is a good redshift indicator \citep{Ric02b},
is shifted entirely beyond the BOSS spectral coverage near $z\sim2.5$.
To measure quasar redshifts above $z\sim2.5$ we rely on combinations of the
C\,III\,$\lambda$1908, C\,IV\,$\lambda$1549, and
Ly\,$\alpha$\,$\lambda$1215 lines.
In addition to being broad, the centroid of each line may be biased---C\,III]
is often blended with Si\,III], Al\,III and Fe\,III
complexes, C\,IV can be shifted from the systemic
redshift by strong outflows, and Ly\,$\alpha$ is often affected by Ly\,$\alpha$
Forest absorption and is blended with NV \citep[e.g.,][]{Van01,Ric02b,Ric11}.

To ameliorate these issues, when the BOSS pipeline\footnote{The 
BOSS pipeline is described in \citet{DR8}} identifies a quasar spectrum
as having an  Mg\,II, C\,III] or C\,IV line that is within the BOSS wavelength
coverage, we default to the redshift from that line offset using the
prescription of \citet{Hew10}.
For spectra with no such lines, we adopt the BOSS pipeline redshift.
The BOSS collaboration is visually inspecting all quasar spectra to check the
pipeline redshifts. When a pipeline redshift conflicts with the visual
redshift (438 objects) we adopt the human-corrected redshift.
As we mainly analyze clustering on scales that correspond to velocities
that are larger than typical quasar broad lines, altering redshifts by a
small amount does not strongly affect our results.
Appendix \ref{sec:zerror} discusses redshift errors further.

\subsection{Quasar luminosities and k-corrections}

Fig.~\ref{fig:magz} plots the conditional magnitude distribution of our
sample, compared to the characteristic luminosity of quasars at that
redshift.  We correct all magnitudes to $z=2$ using the $k$-corrections
derived by \citet{Ric06}.
The characteristic luminosity---where the luminosity function changes 
slope---from \citet{Cro04}, as modified by \citet{Croton09}, is
\begin{equation}
  M_{i,\star}(z) = -21.61 - 2.5\left(k_1 z + k_2 z^2\right) - 0.71
  \quad ,
\end{equation}
where $k_1=1.39$ and $k_2=-0.29$ for $z<3$ and $k_1=1.22$ and
$k_2=-0.23$ for $z\ge 3$.  We have converted from the $b_J$ band used
by \citet{Cro04} to the $i$-band ($k$-corrected to $z=2$)
using $M_i(z=2)=M_{b_J}-0.71$ \citep{Ric06}.
Note that in the range $2.2<z<2.8$, which will be the focus of this paper,
we are able to probe 1--2 magnitudes fainter than the
characteristic magnitude.

\begin{table}
\begin{center}    
\begin{tabular}{ccccr}
\hline
Sample & Name & Redshift & Magnitude & $N_{\rm qso}$\hphantom{t} \\
\hline
1 & All    & $2.2<z<2.8$ & $[-50.0,-10.0]$ & 27,129 \\
2 & Bright & $2.2<z<2.8$ & $[-50.0,-25.8]$ & 13,564 \\
3 & Dim    & $2.2<z<2.8$ & $[-25.8,-10.0]$ & 13,564 \\
4 & Fid    & $2.2<z<2.8$ & $[-27.0,-25.0]$ & 19,111 \\
5 & LoZ    & $2.2<z<2.4$ & $[-27.0,-25.0]$ &  8,835 \\
6 & HiZ    & $2.4<z<2.8$ & $[-27.0,-25.0]$ &  9,977 \\
\hline
\end{tabular}
\end{center}
\caption{\label{tab:summary} A summary of the quasar samples we consider.
The columns list the sample number and name, redshift and magnitude ranges,
and the number of quasars.  Magnitudes are $k$-corrected to $z=2$.}
\label{tab:samples}
\end{table}

Using SDSS broad-band colours to derive $k$-corrections for our sample is
problematic.  Most BOSS quasars at redshift $z\sim3$ are near the flux limit
of SDSS imaging, and so they have noisier colours than for previous SDSS
prescriptions at brighter limits \citep[e.g.][]{Ric06}.
Deriving full $k$-corrections at $z\sim3$ as a function of flux, colour and
redshift is beyond the scope of this paper
(but see McGreer et al.~2012, in preparation). 
Precise $k$-corrections will require proper modeling of subtle changes due to,
e.g., the Baldwin Effect, the presence of complex iron emission crossing
through the $i$-band, and the movement of broad lines---which can be offset
due to luminosity-and-redshift-dependent winds and absorption features---across
the SDSS filter set.
For the redshifts on which we focus in this paper ($2.2<z<2.8$) only the
C\,III\,$\lambda$1908 line enters the $i$-band.
Fortuitously, this complex does not shift much with luminosity, which reduces
any flux dependence to the $k$-correction for our sample.

\section{Clustering}
\label{sec:clustering}

All of our clustering measurements are performed in configuration, rather
than Fourier, space.  For rare objects, where shot-noise is an important
or dominant piece of the error budget, the configuration space estimators
have the advantage of more nearly independent errors.  They also deal
well with irregularly-sampled geometries, such as we have for our sample.
We shall compute both the real- and redshift-space correlation functions,
using the \citet{LanSza93} estimator, with a density of random points $50$
times the density of quasars.

\subsection{The random catalog}

As discussed in \S\ref{sec:datamask} we use {\sc Mangle\/} \citep{Mangle}
to track the angular completeness of the survey.
Angular positions for the random points, modulated by the angular completeness
of the survey, were obtained from the {\sc Mangle\/} program {\tt ransack}.
To assign redshifts to the random catalog we tried three methods, which
yielded almost identical results.  The first was to assign to each point a
redshift drawn at random from the data.  While this method would
produce artificial structure in the redshift distribution of the random
points for a small survey, due to sample variance, the wide angular coverage
of the BOSS survey ensures this method performs well.
We also tried fitting splines to the histogram of the quasar redshifts and
the cumulative histogram of the redshifts and using those splines to generate
random redshifts.  The results were insensitive to using the histogram or
cumulative histogram, to the number of spline points and to the type of spline
used.  For the results presented below we use the first method.

\subsection{Fiber collisions and small-scale clustering}

We cannot obtain spectroscopic data for a few percent of quasars due to
fiber collisions---no two BOSS fibers can be placed closer than $62''$
on a specific plate.  At $z\simeq 2.5$ the $62''$ exclusion
corresponds to $1.26\,h^{-1}$Mpc (comoving).
Where possible we obtain redshifts for the collided quasars in regions where
plates overlap. We account for the remaining exclusions by restricting 
our analyses to relatively large scales and by up-weighting quasar-quasar pairs 
with separations smaller than $62''$.  The upweighting is
derived by comparing the angular correlation function of all targets 
with that of the quasars for which we obtained redshifts
\citep{Haw03,Li06,Ros07,Whi11}. This ratio is close to unity above $62''$
but drops to about two-thirds below $62''$.
The number of pairs for which this correction is appreciable is quite small,
and the impact of this correction is much less than
$1\sigma$ even on the smallest scales.
If fiber-collided quasars preferentially live in regions of higher-than-average 
density the large-scale clustering would be affected by fiber collisions and
would not be properly corrected by our weighting procedure.
The efficiency of the tiling algorithm, the prioritization of quasar targets
over galaxies and the depth of the survey combine to make fiber collisions a
very small effect on our analysis \citep[see also][]{Ros09}.
%A detailed analysis of small scale clustering could be done by
%cross-correlating the imaging data with the spectroscopic sample, but we
%will not attempt this here.

\begin{figure}
\begin{center}
\resizebox{3.4in}{!}{\includegraphics{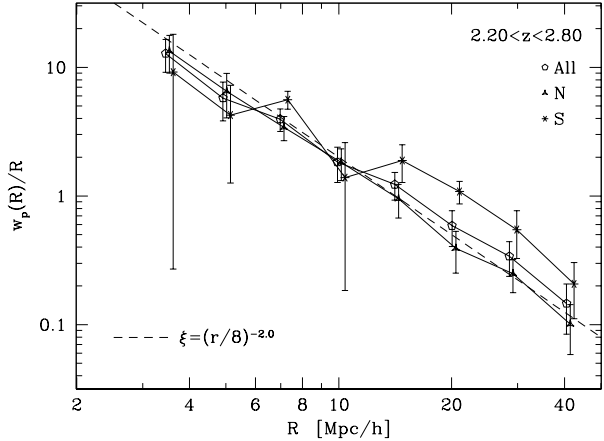}}
\end{center}
\caption{The projected correlation function split by hemisphere (or
Galactic latitude), compared to the fiducial sample.
The dashed line corresponds to the projected correlation function for
a real-space correlation function with $r_0=8\,h^{-1}$Mpc and a power-law
slope of $-2$ to guide the eye.  Note the weakly significant excess
power at large scales for the south-only sample (see text).}
\label{fig:wp_northsouth}
\end{figure}

\begin{figure}
\begin{center}
\resizebox{3.4in}{!}{\includegraphics{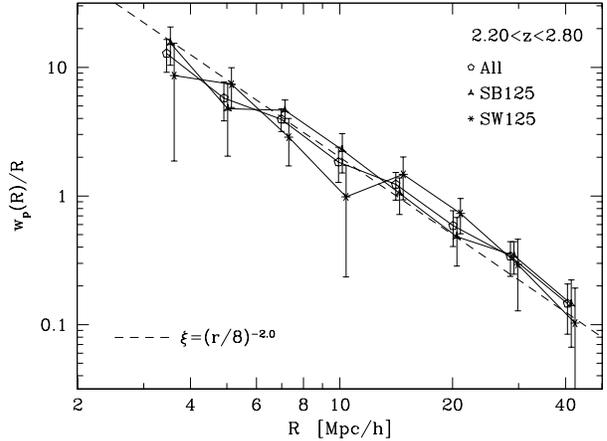}}
\end{center}
\caption{The projected correlation function split by whether the median seeing
in $g$ band in a sector is better than $1.25''$ (SB125) or worse than $1.25''$
(SW125) compared to the fiducial sample.  The dashed line corresponds to the
projected correlation function for a real-space correlation function with
$r_0=8\,h^{-1}$Mpc and a power-law slope of $-2$ to guide the eye.  There
is no statistically significant difference between the two halves of the data.
This is typical of the other jackknife tests we have performed.}
\label{fig:wp_seeing}
\end{figure}

\subsection{Tests of systematics}

We have performed numerous jackknife tests to check whether our results
are robust to possible systematics.  Specifically we have investigated whether
our results are stable to cuts on targeting, at what point in the survey the
plates were drilled and what targeting algorithm was used, sector completeness,
Galactic latitude and hemisphere (Fig.~\ref{fig:wp_northsouth}),
extinction in the $g$-band, target areal density,
stellar density, raw $i$-band magnitude
(which is a proxy for signal-to-noise ratio), sky brightness,
$g$-band seeing (Fig.~\ref{fig:wp_seeing})
and selection threshold.
In all cases but one we see no evidence for a statistically significant
systematic effect.  The exception is that there is weak evidence that the
large-scale clustering of quasars in the South Galactic Cap is stronger than
that in the North Galactic Cap.  It will require more data to determine
whether this is a statistical fluctuation or a significant difference---and, of
course, we conducted {\em eleven}, not one, different tests of systematics.
When eleven (independent) trials are performed the likelihood of a 2$\sigma$
detection is quite high ($\sim40$ per cent instead of $\sim5$ per cent for a
single trial).
In future a quasar catalog with good photometric redshifts could help with
some of these issues.
In addition BOSS will continue to obtain quasar data until 2014, so we defer
a more detailed investigation of geographical discrepancies to a future
publication.

\begin{figure*}
\begin{center}
\resizebox{6in}{!}{\includegraphics{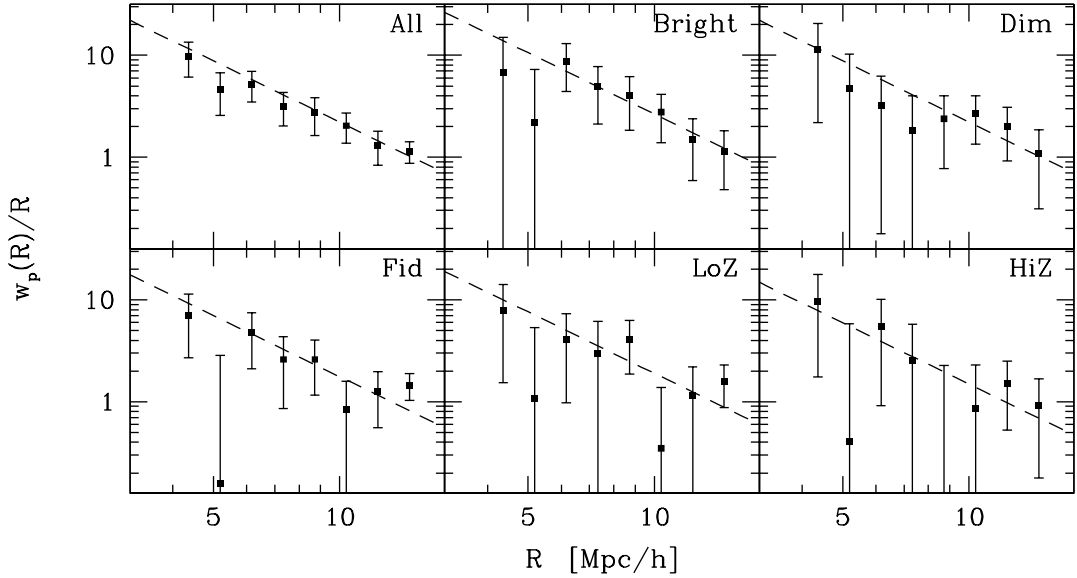}}
\end{center}
\caption{The projected correlation functions, $w_p(R)$, for the six samples
considered in this paper (Table \protect\ref{tab:samples}).
The error bars are the square roots of the diagonal elements of the covariance
matrices, as determined by bootstrap resampling (see text).
The dashed lines show the best fit power-laws with slope $-2$
(see Table \ref{tab:results}).}
\label{fig:allwp}
\end{figure*}

\subsection{Clustering results}

We have insufficient sensitivity to measure the angular dependence of the
redshift-space clustering induced by redshift space distortions for our highly
biased quasars. Therefore, we only quote redshift-space results from the 
angle-averaged correlation function, which we denote $\xi(s)$ at redshift
space separation $s$.
Real-space clustering is constrained by the projected correlation function
\begin{equation}
  w_p(R) \equiv \int dZ\ \xi\left(R,Z\right)
\label{eqn:wpdef}
\end{equation}
avoiding the need to model redshift space distortions and mitigating any
effects of redshift errors.  We truncate the integral over the line-of-sight
separation, $Z$, to $\pm 50\,h^{-1}$Mpc.
This value represents a trade-off between the goal of fully integrating out
the effects of redshift space distortions and the disadvantages of introducing
noise from only weakly correlated structures along the line-of-sight and
mixing a wide range of 3D scales into a single $R$ bin.
By $50\,h^{-1}$Mpc the effects of redshift space distortions are negligible,
and the truncation has only a modest effect on our largest scale point.
However this truncation must be kept in mind when precise modeling of the data
at the largest $R$ is important (see below).

\begin{table*}
\begin{center}
\begin{tabular}{c|rrrrrrrr}
     $R$ &   4.36  &   5.19  &   6.17  &   7.34  &   8.72  &  10.37 
 &  12.34  &  14.67 \\ \hline
 $w_p$ &  42.87  &  24.18  &  32.12  &  23.21  &  23.98  &  21.13 
 &  16.25  &  16.80 \\ \hline
$\sigma$ &  17.08  &  10.89  &  10.77  &   8.36  &   9.85  &   7.12 
 &   6.16  &   4.08 \\ \hline
  4.36  &  1.000  &  0.459  &  0.156  &  0.125  &  0.187  &  0.154 
 & -0.198  & -0.146 \\
  5.19  &     --  &  1.000  &  0.210  &  0.094  &  0.261  &  0.032 
 & -0.256  & -0.019 \\
  6.17  &     --  &     --  &  1.000  &  0.341  & -0.029  &  0.118 
 &  0.168  &  0.173 \\
  7.34  &     --  &     --  &     --  &  1.000  &  0.069  &  0.404 
 & -0.113  & -0.076 \\
  8.72  &     --  &     --  &     --  &     --  &  1.000  &  0.004 
 &  0.053  &  0.091 \\
 10.37  &     --  &     --  &     --  &     --  &     --  &  1.000 
 & -0.107  & -0.119 \\
 12.34  &     --  &     --  &     --  &     --  &     --  &     -- 
 &  1.000  &  0.251 \\
 14.67  &     --  &     --  &     --  &     --  &     --  &     -- 
 &     --  &  1.000 
\end{tabular}
\end{center}
\caption{The $w_p$ data for sample \#1, ``All'', (the largest data set).
The first 3 rows list the transverse separation, $R$, $w_p$ and its error
(all in $h^{-1}$Mpc).  The remainder of the table presents the correlation
coefficients as estimated from the covariance matrix computed using
bootstrapping, as described in the text.  These values are plotted as the
greyscale image in Fig.~\protect\ref{fig:wpcov100}.}
\label{tab:wpX.100}
\end{table*}

\begin{figure*}
\begin{center}
\resizebox{6in}{!}{\includegraphics{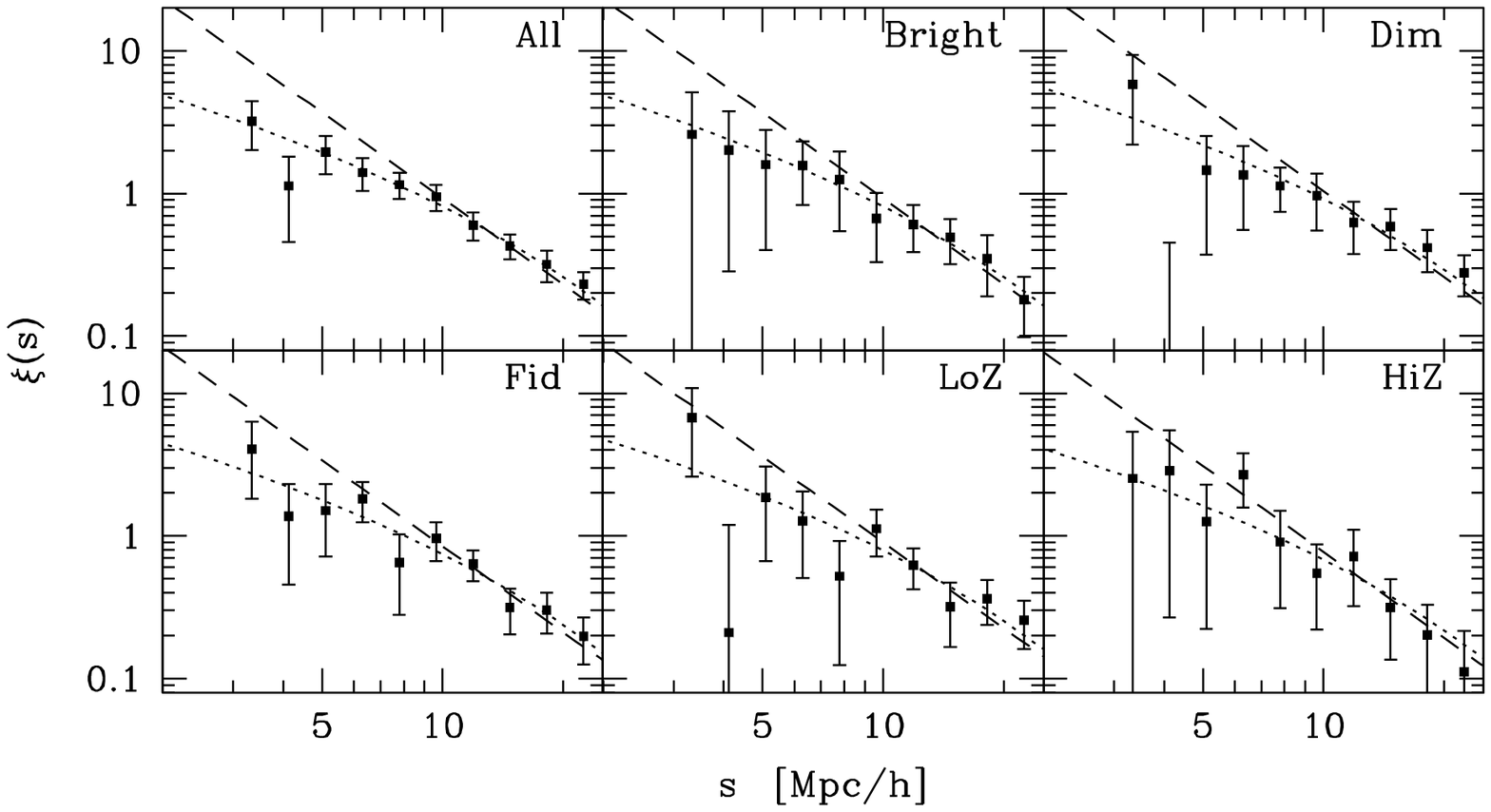}}
\end{center}
\caption{The redshift-space correlation function, $\xi(s)$, for the six samples
considered in this paper.  The error bars are the square roots of the diagonal
elements of the covariance matrices, as determined by bootstrap resampling
(see text).
The dashed lines show the best fit power-laws with slope $-2$
(see Table \ref{tab:results}) while the dotted lines show the power-law once
redshift errors are taken into account (see Appendix \protect\ref{sec:zerror}).}
\label{fig:allxi}
\end{figure*}

We divide our quasar sample into bins of redshift over which the bias and
mass correlation function are evolving strongly.  Fortuitously, on the
scales of relevance the effects approximately cancel, i.e.~the clustering
amplitude stays approximately constant.
The redshift-bin-averaged $\xi$ can be approximated as a measurement of
$\xi$ evaluated at an effective redshift, $z_{\rm eff}$:
\begin{equation}
  z_{\rm eff} =
  \frac{\int dz\ (dN/dz)^2(H/\chi^2)\ z}{\int dz\ (dN/dz)^2(H/\chi^2)}
\label{eqn:zeff}
\end{equation}
so that
\begin{equation}
  \xi(s,z_{\rm eff}) \simeq \langle\xi(s)\rangle =
  \frac{\int dz\ (dN/dz)^2(H/\chi^2)\xi(s,z)}{\int dz\ (dN/dz)^2(H/\chi^2)}
\end{equation}
where $dN/dz$ is the redshift distribution of the sample, $H$ is the Hubble
parameter at redshift $z$ and $\chi$ is the comoving angular diameter distance
to redshift $z$ \citep{Mat97,WhiMarCoh08}.
Assuming passive evolution, constant halo mass, or constant bias leads to 
differences between $\xi(z_{\rm eff})$ and $\langle\xi\rangle$ that are far
smaller than our observational errors.

\begin{figure}
\begin{center}
\resizebox{3.4in}{!}{\includegraphics{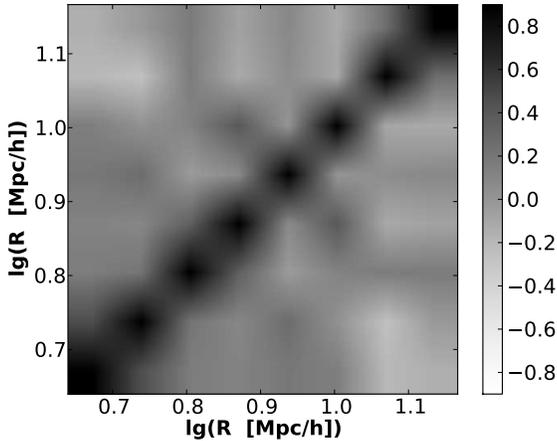}}
\end{center}
\caption{The correlation matrix for the projected correlation function,
$w_p(R)$, of quasars with $2.2<z<2.8$ and no cuts on magnitude (i.e.~sample
\#1).}
\label{fig:wpcov100}
\end{figure}

\begin{figure}
\begin{center}
\resizebox{3.4in}{!}{\includegraphics{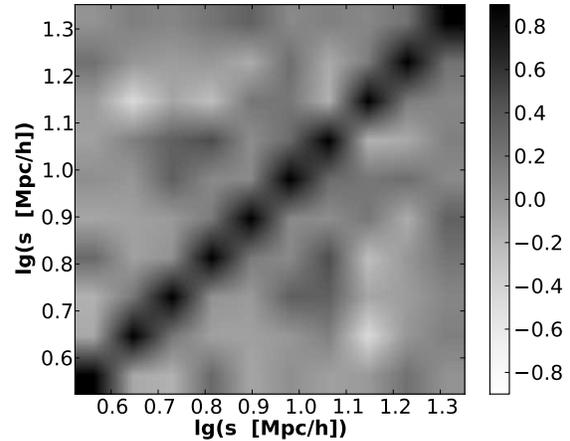}}
\end{center}
\caption{The correlation matrix for the redshift-space correlation function
$\xi(s)$, of quasars with $2.2<z<2.8$ and no cuts on magnitude (i.e.~sample
\#1).}
\label{fig:xicov100}
\end{figure}

\begin{figure}
\begin{center}
\resizebox{3.4in}{!}{\includegraphics{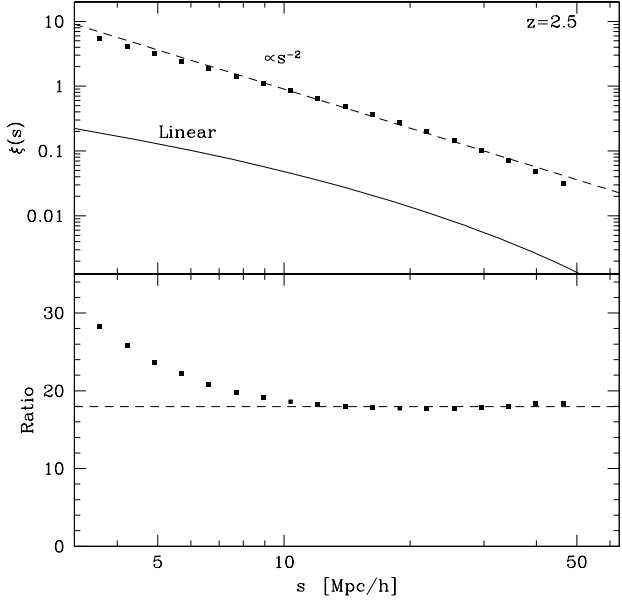}}
\end{center}
\caption{The clustering of a sample of halos spanning an octave (i.e.~factor
of 2) in mass centered on $2\times 10^{12}\,h^{-1}M_\odot$.
This sample has a narrow range of halo masses with a clear ``characteristic''
mass while also containing enough halos to enable a precision estimate of the
correlation function.
The solid line shows the linear-theory, real-space correlation function for
our fiducial cosmology while the dashed line is a power-law with slope $-2$.
The points are computed from halos at $z=2.5$ from two simulations of $3000^3$
particles ($m=6\times 10^{10}\,h^{-1}M_\odot$) each in a $2.75\,h^{-1}$Gpc box
run using the code described in \protect\citet{TreePM}.  The halos are found
using the friends-of-friends method \citep{DEFW} with a linking length of
0.168 times the mean inter-particle separation.
In the upper panel the points show the mean of the angle-averaged,
redshift-space correlation functions computed from the periodic boxes in the
distant observer approximation.
The lower panel shows the ratio of the two, which becomes constant at large
scales.
The horizontal dashed line is a fit to the asymptote between
$20<s<40\,h^{-1}$Mpc.  Assuming $\xi(s)=[b^2+(2/3)bf+f^2/5]\xi_{\rm lin}$
\protect\citep{Kai87} with $f\equiv [\Omega_m(z)]^{0.56}$ gives $b\simeq 3.9$,
in good agreement with the value inferred from the real-space clustering.
The slope and amplitude of the power-law piece of $\xi(s)$, and how low in
$s$ it extends, varies with the particular halo subsample chosen, but the
scale-independence of the bias at large scales is generic.}
\label{fig:halo_xi}
\end{figure}

\begin{figure}
\begin{center}
\resizebox{3.4in}{!}{\includegraphics{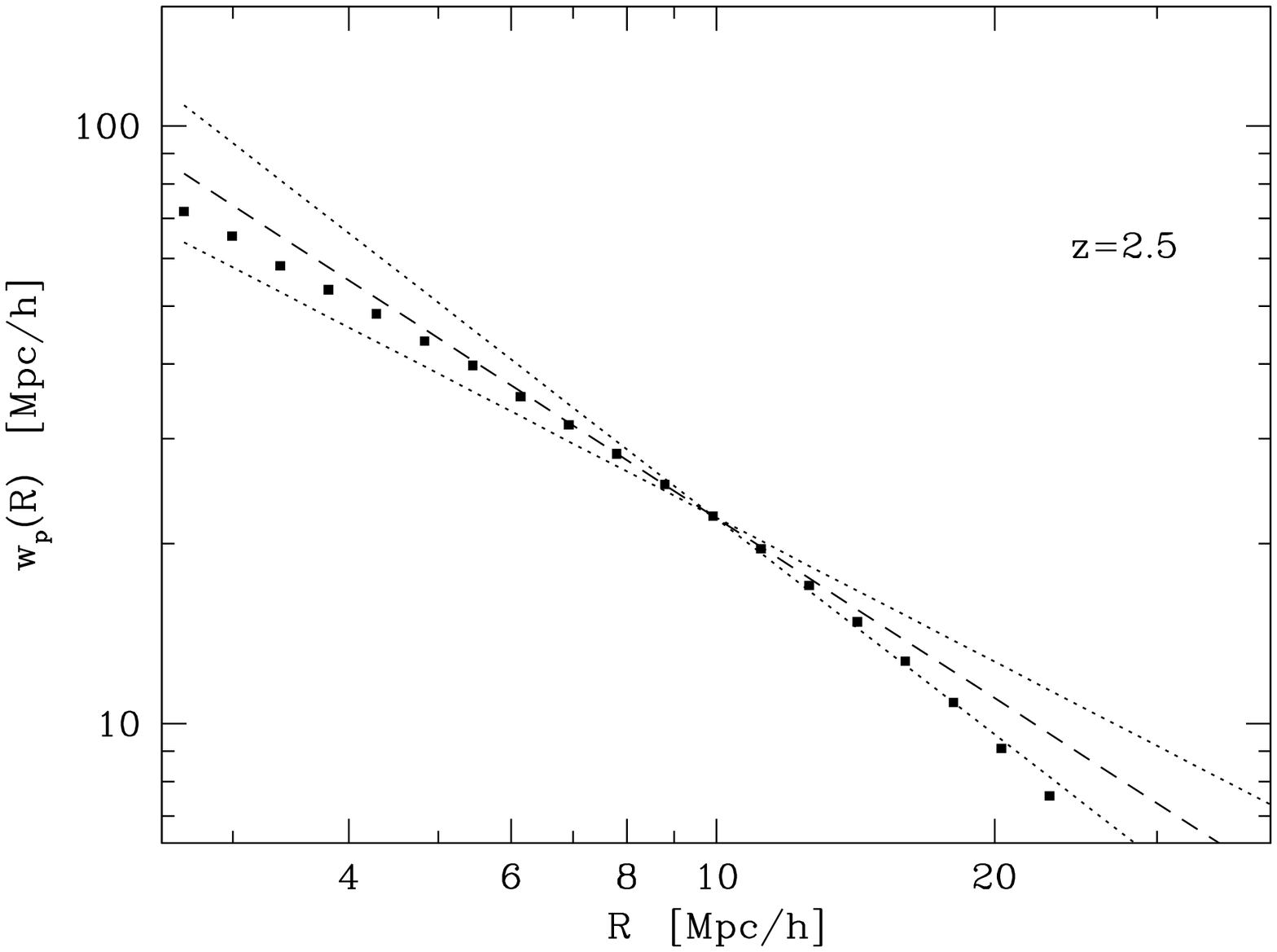}}
\end{center}
\caption{The projected correlation function for the same sample of halos
described in Fig.~\protect\ref{fig:halo_xi}.  The dotted and dashed lines
correspond to power-laws of slope $-0.8$, $-1$ and $-1.2$,
arbitrarily normalized to the data at $R\simeq 10\,h^{-1}$Mpc.
We imposed an upper limit of $\pm 50\,h^{-1}$Mpc on the line-of-sight
separation in Eq.~(\ref{eqn:wpdef}).
Changing the mix of halos in the sample can change the amplitude and slope
of $w_p$, but the break is a generic feature.}
\label{fig:halo_wp}
\end{figure}

Results for each of the samples in Table \ref{tab:samples} are shown in
Figs.~\ref{fig:allwp} and \ref{fig:allxi}, and the values for the full
sample, with $2.2<z<2.8$ and no cuts on magnitude, are given in
Table \ref{tab:wpX.100}.
Both the real- and redshift-space clustering are fit well by a model with
an underlying power-law correlation function, once the effects of projection
and redshift errors are taken into account.  As it provides a good fit to the
data, and for consistency with earlier work \citep[e.g.][]{Mye06,Ros09,She09},
we shall show lines in the figures and provide fits in the tables assuming a
power-law slope of $r^{-2}$ for the real-space correlation function.
The actual slope of the correlation function is poorly determined by the
projected correlation function data.  Power-law indices from $-1$ to $-2.6$
are viable.  The best-fit slope is quite shallow, near $-1.2$.
The value of $w_p(R)$ at $R\simeq 9\,h^{-1}$Mpc for the best-fit model is
almost independent of the assumed slope.

We estimate the covariance matrix of our measurements by bootstrap resampling
\citep[e.g.][]{Efron}.
We divide the survey into angular regions specified by {\sc HealPix\/}
pixels \citep{Healpix} with $N_{\rm side}=4$ (i.e.~approximately $15^\circ$
on a side).
Pixels which contain fewer random points than
two-thirds of the mean are merged with higher occupancy pixels
to ensure pixels have similar weight. We then estimate both the 
mean and covariance matrix by bootstrap
resampling pair counts from a random draw of pixels (with
replacement).

The bootstrap-determined correlation matrices for the full $2.2<z<2.8$ sample
for $w_p$ and $\xi$ are shown in Figs.~\ref{fig:wpcov100} and \ref{fig:xicov100}
respectively.  Note that the matrices are diagonal-dominated as might be
expected for shot-noise limited measurements---extending to larger scales we
see larger correlations between the bins, most notably in $w_p$.
As we begin to restrict the quasar sample in redshift or luminosity, and the
number of objects becomes smaller, the covariance matrices and their
inverses can become increasingly noisy.  We have several options at
this point---with the most desirable being a reduction in the number of
degrees of freedom (i.e.~data-compression) so that we can apply bootstrap
resampling to obtain a converged covariance matrix, or error on a summary
statistic (or statistics).
The simplest approach would be to reduce the number of bins so that we have
fewer, better constrained points.  However this is not optimal for our
purposes.  We describe below a different approach based on the sparsity of our
sample and the nature of the clustering.

To begin we note that the correlation functions are fit well
($\chi^2=4.5$ for 7 degrees of freedom for $w_p$, and $\chi^2=12$
for 9 degrees of freedom for $\xi$)
by power-laws over the range where our constraints are tightest (as expected
if quasars are hosted by massive halos, see Fig.~\ref{fig:halo_xi}).
We adopt a two parameter model for both $\xi(s)$ and $w_p$ of the form
$\xi=(s_0/s)^\gamma$ and
\begin{equation}
  \frac{w_{p}(R)}{R} =
  \frac{\sqrt{\pi}\,\Gamma[(\gamma-1)/2]}{\Gamma[\gamma/2]} 
  \left(\frac{r_0}{R}\right)^{\gamma} \,\,,
\label{eqn:wp-xi}
\end{equation}
which corresponds to a 3D correlation of the form
$\xi(r) = (r_0/r)^{\gamma}$ integrated to $\pm\infty$ along the line-of-sight.
For $\gamma=2$ the prefactor is $\pi$ and we shall fix $\gamma=2$ throughout so
that we have one remaining degree of freedom.
We fit these models to the measured correlations over the range
$3<s<25\,h^{-1}$Mpc and $4<R<16\,h^{-1}$Mpc.
The non-linear, scale-dependent bias of dark matter halos makes their
correlation functions close to a power-law on Mpc scales
(see Figs.~\ref{fig:halo_xi} and \ref{fig:halo_wp})
but at larger scales the bias becomes scale-independent and the correlation
functions drop more quickly than the power-law extrapolation would suggest.
In addition, on larger scales sample variance becomes increasingly important
and the radial bins become increasingly correlated (especially for $w_p$).
For these reasons we limit the fitting range as indicated
\citep[see also][]{Cro05}.
With more data and a numerical model for the covariance matrix we could extend
the range of the measurement and tighten the constraints on quasar models.

For a power-law correlation function of fixed slope each point provides an
estimate of the correlation length, $s_0$, since
$s_0^\gamma=s_i^\gamma\xi(s_i)$.
The number of pairs in a bin of fixed $\Delta\log s$ is
$N_{\rm pair}\propto (1+\xi)s^3$ and if shot-noise dominates the bins are
independent and the fractional error on $1+\xi(s)$ in each bin scales as
$N_{\rm pair}^{-1/2}$.
Thus, assuming logarithmic bins and that shot-noise dominates we can average
the estimates with inverse variance weights so that
\begin{equation}
  s_0 = \left(\sum_i s_i^\gamma \xi(s_i) w_i\middle/
         \sum_i w_i\right)^{1/\gamma}
\label{eqn:s0fit}
\end{equation}
where the weights are
\begin{equation}
  w_i^{-1} \propto \left[1+\left(\frac{s_0'}{s_i}\right)^\gamma\right]
           s_i^{2\gamma-3} \qquad .
\end{equation}
The values of $w_i$ depend on an estimate, $s_0'$, for $s_0$, and reduce to
\begin{equation}
  w_i \propto \frac{s_i}{s_i^2+s_0'^2}
  \qquad {\rm for}\ \gamma=2 \quad .
\end{equation}
Most of the weight in the fit is produced by $s_i\sim s_0'$, with the weight
scaling as $s^{3-\gamma}$ for small $s$ and $s^{3-2\gamma}$ for large $s$.
Since the bins are assumed to be logarithmically spaced, for $\gamma\approx 2$
this suppression is quite rapid in both directions, reflecting the paucity of
pairs at small $s$ and the weakness of the correlation signal at large $s$.
If the correlation function continued as a power-law to large $s$ we could
tighten our constraints by extending the fitting range, but it is at these
larger scales where deviations from a power-law behavior are most expected,
and where the correlations between measurements in adjacent radial bins act
to weaken the constraints.

Unfortunately, it is difficult to find a precise redshift for quasars in
the range $2<z<3$ from optical spectroscopy, so our correlation function
is smeared by redshift errors which reduce the small-scale clustering
signal (the dotted lines in Fig.~\ref{fig:allxi}; Appendix \ref{sec:zerror}
discusses the impact of redshift uncertainties further).
We include redshift errors in our model through a
multiplicative factor, $\xi(s)\to F(s)\xi(s)$, derived in
Appendix \ref{sec:zerror}.  The best fit $s_0$ can be derived from a
generalization of Eq.~(\ref{eqn:s0fit}) which replaces the weights with
$w_i\to F^2(s_i)w_i$, multiplies the $\xi(s_i)$ in the weights by $F(s_i)$
and divides each term in the numerator of Eq.~(\ref{eqn:s0fit}) by $F(s_i)$.

For a given estimate, $s_0'$, the optimal estimate of $s_0$ can be written
either as a weighted sum of the $\xi_i$ bins or directly in terms of the
pair counts themselves.  One can iterate this estimator, in which case the
error properties of $s_0$ become more complex and are best handled by
a bootstrap procedure.
We generate $s_0$ for each bootstrap sample, with the iterative weighting
scheme above starting from $s_0'=0$, and use the standard deviation for our
estimate of uncertainty.
This uncertainty estimate does not include the additional contribution from
our uncertainty in the redshift error correction.  A change in redshift
error ($\sigma_z$) of $\pm 25$ per cent moves the best-fit $s_0$ by $1\sigma$,
which can be considered an additional systematic error on the fit.
While a 25 per cent uncertainty on the redshift error is consistent with 
Appendix \ref{sec:zerror}, we find that reasonable fits to $\xi(s)$ can be
obtained for a wide range of redshift error due to a degeneracy between the
assumed redshift error and the amplitude (and slope) of the underlying
correlation function.
Because of this additional uncertainty, our strongest cosmological constraints
will come from the projected correlation function, to which we now turn.

A similar estimator can be used for the real-space correlation length, $r_0$,
under the same assumptions.  The weights in this case become
$w_i^{-1}\propto (2Z_{\rm max}+w_{p\,i})R_i^{2\gamma-4}$ if the integration in
Eq.~\ref{eqn:wpdef} extends from $-Z_{\rm max}$ to $Z_{\rm max}$.
For $\gamma\approx 2$ the weights are nearly constant for all the samples
we consider for all $R$ of interest.
As for the case of the redshift errors and $\xi(s)$, we can account for the
effects of finite $Z_{\rm max}$ on Eq.~\ref{eqn:wp-xi} by modifying the
weights to $w_i\to F^2(R_i)w_i$ and dividing each term in the numerator by
$F(R_i)$, where\footnote{Assuming that $Z_{\rm max}$ is large enough that
fingers-of-god are correctly included and $b\gg 1$ so that the anisotropy
due to redshift-space distortions is small.}
$F(R)=(2/\pi)\arctan(Z_{\rm max}/R)$ for $\gamma=2$.

\begin{figure}
\begin{center}
\resizebox{3.4in}{!}{\includegraphics{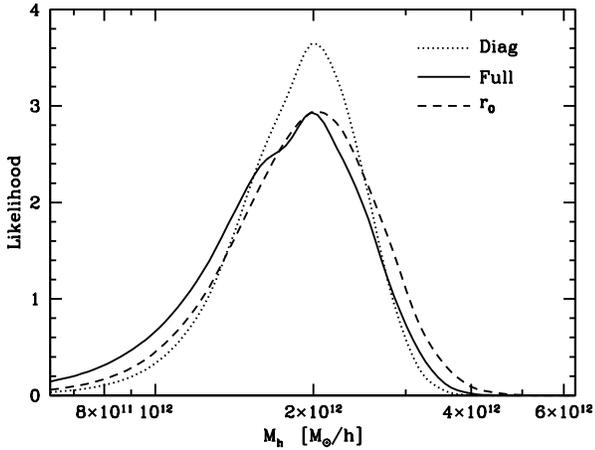}}
\end{center}
\caption{A comparison of fits to the real-space clustering data, using a
variety of approximations.  The model in each case is the clustering of a
sample of halos covering one octave in mass, centered on $M_h$.
The likelihood of the central mass is computed by fitting to the $w_p$
measured for sample 1 using the full covariance matrix determined by bootstrap
(solid), a diagonal covariance matrix (dotted) or the value of $r_0$ determined
as described in the text (dashed).  For this model the central mass has an
error of $0.5\times 10^{12}\,h^{-1}M_\odot$.}
\label{fig:likelihood_compare}
\end{figure}

The clustering strength derived from this procedure is a statistically valid
summary of the data under the assumption that a power-law provides a good
fit to $w_p$.  However it does not need to be an optimal compression of the
available information --- if the data are sufficiently informative a better
constraint on the clustering amplitude could be obtained by fitting all of
the data.
For the full sample (\#1 in Table \ref{tab:samples}), where we have a
reasonably converged estimate for the covariance matrix, we can compare the
different methods.
In this case, the likelihood derived from the $r_0$ determined as above is
very similar to that derived from the full covariance matrix (or the diagonal
elements) indicating that in our case our estimate of $r_0$ does provide an
almost exhaustive summary of the constraints available
(Fig.~\ref{fig:likelihood_compare}).
This will be even more the case for samples with more shot-noise.
Our simple estimator is the preferred approach for quoting clustering
measurements and errors on sparse samples where estimating a full covariance
matrix is not feasible.

In addition to our estimates of $r_0$ and $s_0$ we also provide another
summary statistic \citep[motivated by][]{Cro05,Ang08,Ros09},
\begin{equation}
  \bar{\xi} \equiv \frac{3}{s_{\rm max}^3-s_{\rm min}^3}
    \int_{s_{\rm min}}^{s_{\rm max}} s^2 ds\ \xi(s)
\label{eqn:xibardef}
\end{equation}
with $s_{\rm min}=5\,h^{-1}$Mpc and $s_{\rm max}=20\,h^{-1}$Mpc.
For $\xi(s)=(s_0/s)^2$ Eq.~(\ref{eqn:xibardef}) becomes
\begin{eqnarray}
  \bar{\xi} &=&
  \frac{3s_0^2}{s_{\rm max}^2+s_{\rm max}s_{\rm min}+s_{\rm min}^2} \\
  &\approx& 3\left(\frac{s_0}{s_{\rm max}}\right)^2
  \left[1-\frac{s_{\rm min}}{s_{\rm max}} + \cdots \right]
  \label{eqn:xibarexpansion}
\end{eqnarray}
where the last step assumes $s_{\rm min}\ll s_{\rm max}$.
We adopt a lower limit, $s_{\rm min}\ne 0$, to mitigate the effect
of redshift errors and scale-dependent bias---Eq.~\ref{eqn:xibarexpansion}
shows that this differs from the $s_{\rm min}$ case by 25 per cent for a
power-law of index $-2$.
With our lower limit the bias inferred from modeling $\bar{\xi}$ using the
\citet{Kai87} prescription
\begin{equation}
  \bar{\xi}(s) \approx \left( b^2 + \frac{2bf}{3} + \frac{f^2}{5}\right)
  \bar{\xi}_{\rm real}
\label{eqn:xiapprox}
\end{equation}
agrees to 1 per cent with that inferred from the real-space clustering for
the simulation results described in
Figs.~\ref{fig:halo_xi} and \ref{fig:halo_wp}.
If $\xi_{\rm lin}$ is used in place of $\xi_{\rm real}$ in
Eq.~\ref{eqn:xiapprox}, the error is also 1 per cent for the case shown in
Figs.~\ref{fig:halo_xi} and \ref{fig:halo_wp}, though it becomes larger
for more biased samples.
We compute $\bar{\xi}$ from the data by assuming $\xi(s)$ can be modeled
as a constant within the 10 bins, spaced equally in log between $s_{\rm min}$
and $s_{\rm max}$.
The values are corrected upwards by 7 per cent to account for the effect
of redshift errors.
To allow easy comparison with earlier work we use the value of
$\bar{\xi}$ to estimate the bias in Table \ref{tab:results}.

\begin{table*}
\begin{center}    
\begin{tabular}{cccccccccc}
\hline
Sample & Redshift & $z_{\rm eff}$ & Magnitude & Median & Mean &
   $r_0$ & $s_0$ & $\bar{\xi}$ & Bias \\ \hline
1 & $2.2<z<2.8$ & $2.39$ & $[-50.0,-10.0]$ & $-25.8$ & $-25.9$ &
   $8.4\pm0.6$ & $9.6\pm0.5$ & $0.52\pm0.06$ & $3.8\pm0.3$ \\
2 & $2.2<z<2.8$ & $2.41$ & $[-50.0,-25.8]$ & $-26.5$ & $-26.6$ &
   $9.2\pm0.9$ & $9.6\pm1.0$ & $0.54\pm0.12$ & $3.9\pm0.5$ \\
3 & $2.2<z<2.8$ & $2.36$ & $[-25.8,-10.0]$ & $-25.2$ & $-25.1$ &
   $8.4\pm1.7$ & $10.2\pm0.8$ & $0.65\pm0.12$ & $4.3\pm0.4$ \\
4 & $2.2<z<2.8$ & $2.39$ & $[-27.0,-25.0]$ & $-25.9$ & $-25.9$ &
   $7.5\pm0.9$ & $9.2\pm0.7$ & $0.48\pm0.06$ & $3.7\pm0.3$ \\
5 & $2.2<z<2.4$ & $2.28$ & $[-27.0,-25.0]$ & $-25.8$ & $-25.9$ &
   $7.8\pm1.3$ & $9.5\pm0.9$ & $0.51\pm0.10$ & $3.7\pm0.4$ \\
6 & $2.4<z<2.8$ & $2.51$ & $[-27.0,-25.0]$ & $-25.9$ & $-25.9$ &
   $6.9\pm1.8$ & $8.8\pm1.1$ & $0.43\pm0.13$ & $3.6\pm0.6$ \\
\hline
\end{tabular}
\end{center}
\caption{A summary of our clustering results.  The samples are as in
Table \ref{tab:samples}.  The effective redshift is computed using
Eq.~\protect\ref{eqn:zeff}.  The ranges, medians and mean magnitudes
refer to absolute, $i$-band magnitude $k$-corrected to $z=2$.  The
correlation lengths are measured in (comoving) $h^{-1}$Mpc.  The
$1\,\sigma$ errors on $r_0$, $s_0$ and $\bar{\xi}$ are from bootstrap
resampling, as described in the text.
The error for $s_0$ does not include the additional uncertainty 
due to the redshift error from the pipeline.  The bias is
estimated from $\bar{\xi}$ using Eqs.~\protect\ref{eqn:xibardef} and
\protect\ref{eqn:xiapprox} with the fitting function of \citet{HaloFit}
for the real-space correlation function of the mass.}
\label{tab:results}
\end{table*}

%\begin{figure}
%\begin{center}
%\resizebox{3in}{!}{\includegraphics{pk.eps}}
%\end{center}
%\caption{The redshift-space power spectrum.}
%\label{fig:pk100}
%\end{figure}

\begin{figure}
\begin{center}
\resizebox{3.4in}{!}{\includegraphics{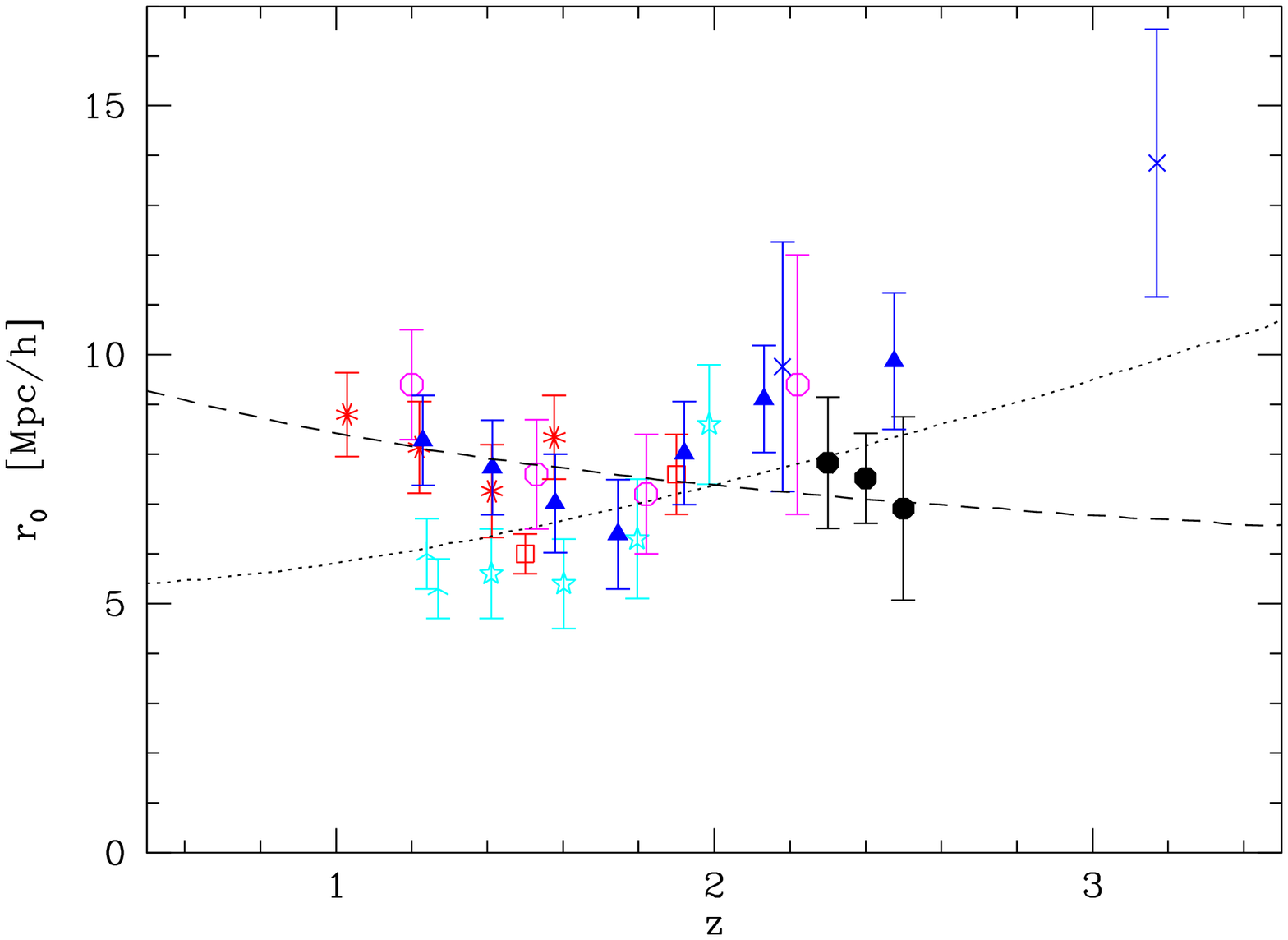}}
\resizebox{3.4in}{!}{\includegraphics{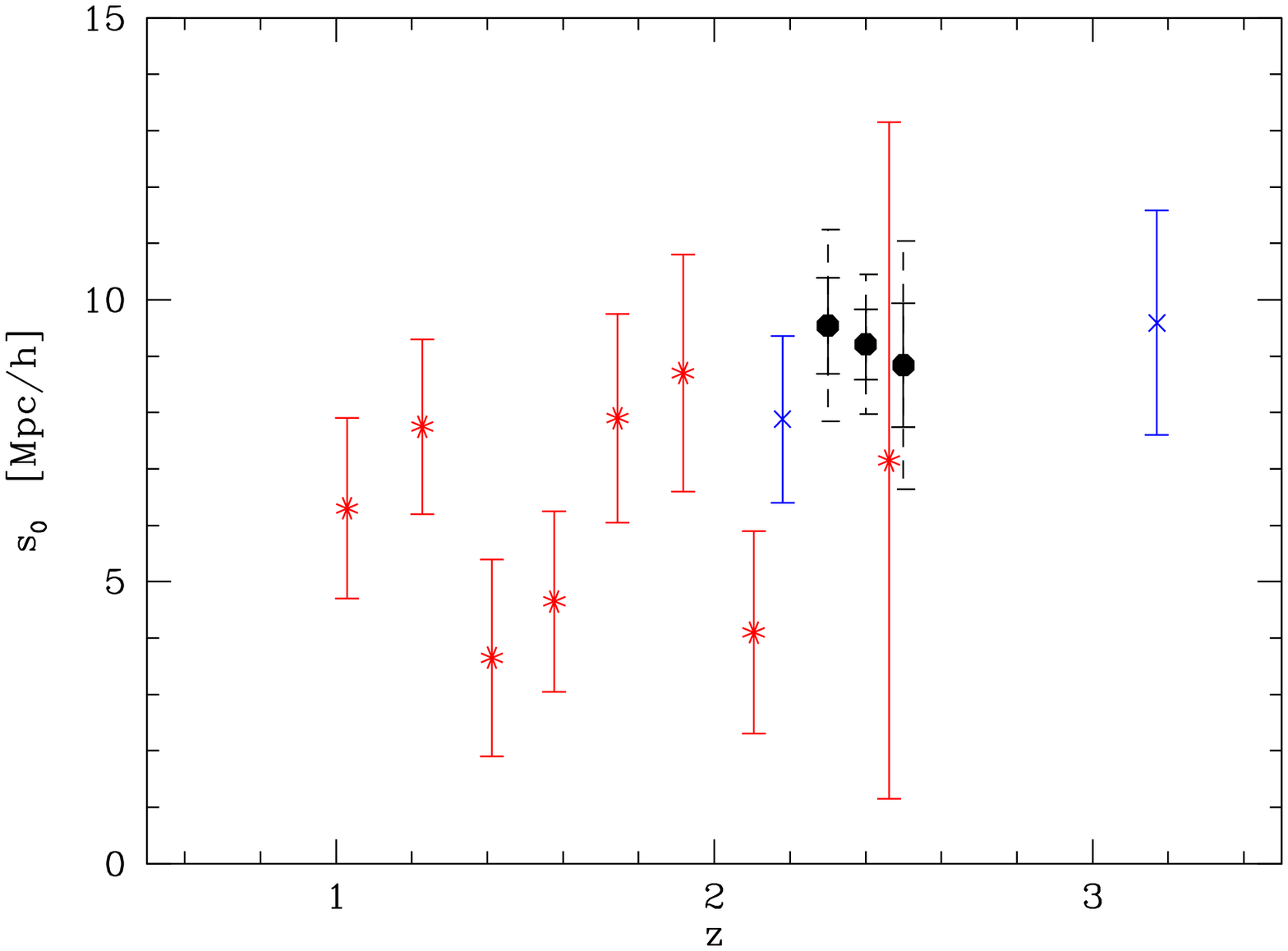}}
\end{center}
\caption{Estimates of the real- and redshift-space correlation lengths for
our last three samples (\#4-\#6, solid circles) and from previous work.
Open squares are from \citet{PorMagNor04}, solid triangles from
\citet[][converted from $\bar{\xi}$ measurements]{Cro04},
open 5-pointed stars from \citet{PorNor06}, open octagons from \citet{Mye06},
8-pointed stars from \citet{Ros09}, crosses from \citet{Shen++09}
and 3-pointed stars from \citet[][obscured and unobscured]{Hic11}.
In the upper panel the dotted line indicates the evolution of $r_0$ for a
sample of halos of a single mass, $M=2\times 10^{12}\,h^{-1}M_\odot$, while
the dashed line shows the evolution of a passively evolving sample with no
mergers \protect\citep{Fry96}.  Both lines are meant for illustrative purposes,
and there has been no attempt to fit to the data.  In the lower panel the
solid error bars plotted exclude the contribution from our uncertainty in the
redshift error, which is significant (see text).  The dashed error bars show
the effect of doubling the errors for our measurements.}
\label{fig:prevwork}
\end{figure}

%[Include cross-correlation of bright/dim samples with the full sample, as
%in \citet{Shen++09}?  The code is all set up to do this, but currently
%it's quite noisy.]
Our results are listed in Table \ref{tab:results} and
compared to previous work in Figure \ref{fig:prevwork}.
We do not detect a luminosity or redshift dependence of the clustering
strength, although our sensitivity to this dependence is weak due to the
limited dynamic range in both variables in our sample.
When comparing to previous work, we do not plot the last 4 points quoted
in Table 2 of the SDSS Data Release 5 \citep[DR5][]{DR5} quasar clustering
analysis of \citet{Ros09}.
Due to the flux-limited quasar selection by the original SDSS, the projected
$w_p$ measurement for these high-$z$ bins are quite noisy \citep{She07,Ros09},
and thus unreliable for any direct comparison to our BOSS measurement.  
We have checked that even if we use the SDSS DR7 data this situation is not
improved for our $2<z<3$ redshift range of interest.

Our results strongly favor the consensus that quasars inhabit
rare and highly biased dark matter halos on the exponential tail of the
mass function.  In the absence of merging we would expect the clustering
of such halos to evolve slowly with time.
For our assumed cosmology our quasar samples have biases in the
range 3.4--4, consistent with early observations and the extrapolation
of previous measurements by \citet{Cro05}.
Estimates in the literature on the typical halo mass for a bright quasar at
comparable redshifts vary wildly, in part due to methodological differences
and the fitting functions assumed.
So, we now turn to how quasars occupy dark matter halos
(see also Appendix \ref{sec:halomodel}).

\section{Interpretation and modeling}
\label{sec:interp}

One of the main goals of studying quasar clustering is to provide information
on the parent dark matter halos hosting luminous quasars.  The large-scale
bias of the quasars provides information on the mean dark matter halo mass;
the small-scale clustering provides information on satellite fractions and
potentially radial profiles within halos.  Unfortunately the space density of
quasars cannot be used directly in such constraints because the quasar duty
cycle (or activity time) is not known.
This is a major difference with studies of e.g.,~galaxy clustering, and has
serious implications for the constraints that can be derived.

Our median quasar has $M_i(z=2)=-26$ and so a bolometric luminosity of
$L_{\rm bol}=2.5\times 10^{39}\,$W \citep{Cro05,Shen++09}.
If this object is radiating at the Eddington limit
($L_{\rm Edd} = 10^{40.1} [M_{\rm bh}/10^{9} M_{\odot}]\,$W),
then the median $M_{\rm bh}$ in our sample is $2\times 10^8\,M_\odot$.
As we shall describe, our data are consistent with host halos having a
characteristic mass of $2\times 10^{12}\,h^{-1}M_\odot$
(Fig.~\ref{fig:likelihood_compare}),
in agreement with earlier work \citep{PorMagNor04,Cro05,PorNor06,Lid06}.
This value is also consistent with estimates from the
$M_{\rm bh}-M_{h}$ relation \citep{Fer02,Fin06},
$M_{h}\simeq (1-3)\times 10^{12}\,h^{-1}M_\odot$,
with the differences arising from different assumptions about the halo
profiles or data sets. 
This result suggests $L/L_{\rm edd} \sim 1$, consistent with the results of 
\citet{Cro05}.

However, we don't expect quasars to inhabit halos of a single mass
\citep[though see][]{Sha11}.
Constraining the possibly complex manner in which quasars occupy halos from
the relatively featureless correlation functions we have access to is
difficult, but the modeling is made easier by a number of facts.
Quasars are rare, their activity times are short and the fraction of binary
quasars is very small\footnote{When the virial radius of the hosting halo
is much smaller than the mean inter-quasar separation, the fraction of
quasar-hosting halos which contain a second quasar scales as
$V_{\rm halo}\bar{n}_Q\xi_V\ll 1$, where $\xi_V$ is the volume averaged
correlation function within the virial radius.} \citep{Hen06a,Mye07b}.
This suggests that most quasars live at the center of their dark matter halos
and the majority of halos host at most one active quasar.
To place constraints on the range of halos in which quasars may be
active we consider two illustrative models described below
(see Appendix \ref{sec:halomodel} for further details and \citealt{DeGraf11}
for a recent discussion in the context of numerical hydrodynamic simulations).
To make interpretation easier we use a quasar sample which is limited at
both the bright and faint ends, i.e.~$-27<M_i<-25$ or
$39.0<{\rm log}_{10}\,L_{\rm bol}<39.8$ (see Table \ref{tab:samples},
sample \#4, ``Fid'').
We use the $r_0$ measurements listed in Table \ref{tab:results} to constrain
the models, though similar results are obtained using the $w_p$ data and its
covariance matrix.
We also find that the best-fitting models below provide a good fit to the
redshift space clustering, assuming our fiducial model for redshift errors.

\subsection{Lognormal model} \label{sec:lognormal}

In this model we assume that quasars of some luminosity, $L$, live in halos
with a lognormal\footnote{Quasars are known to have a relatively high bias
and hence live in halos on the steeply falling tail of the mass function.
The differences between occupation models which include a high-$M$ cut-off
and those that do not is therefore relatively small.} distribution of
masses, centered on a characteristic mass that scales with $L$
(see also Appendix \ref{sec:halomodel}).
Each halo hosts at most one quasar with probability
\begin{equation}
  P(M_h|L) \propto \exp\left[
  -\frac{\left(\ln M_h-\ln M_{\rm cen}(L)\right)^2}{2\sigma^2(L)}\right]
\end{equation}
where the normalization is set by the observed space-density of quasars but
does not matter for the clustering.
We expect that $M_{\rm cen}$ will be larger for more luminous quasars that
are hosted by more massive galaxies.

For illustration we assume that the lognormal form holds for quasars in
the luminosity bin $-27<M_i<-25$, i.e.~that the luminosity dependence of
$M_{\rm cen}$ is weak.  This is a reasonable approximation to many models
(see Appendix \ref{sec:halomodel}) and in particular the type described in
the next section.
We generate quasar samples from halos in the $z=2.4$ output of 4
cosmological simulations, each employing $1500^3$ particles
($m=2\times 10^{10}\,h^{-1}M_\odot$) in a $1\,h^{-1}$Gpc box.
As in Fig.~\ref{fig:halo_xi}, halos are found in the simulations by the
friends-of-friends algorithm with a linking length of 0.168 times the mean
inter-particle separation.
The halos are selected with a lognormal probability centered on a series
of $M_{\rm cen}$.  To test for sensitivity to the width of the distribution
we consider $\sigma=25$, 50 and 100 per cent.  The former is more appropriate
for higher redshift or brighter quasars \citep{WhiMarCoh08,ShaWeiShe10,DeGraf11}
while the latter is roughly expected based on the amount of observed 
scatter in the local $M_{BH}-M_{\rm gal}$ relation.
We average the projected correlation function for each model---with the same
binning and $Z_{\rm max}$ as the data---over the 4 simulations to reduce noise,
and compute a goodness-of-fit.
Using either the full bootstrap covariance matrix or just the diagonal entries,
we find that the best-fit model is good for all choices of $\sigma$.
Not surprisingly, we obtain consistent results if we fit $r_0$ to the average
$w_p$ of the simulations using the procedure of \S\ref{sec:clustering} and then
compare to the value in Table \ref{tab:results}.
Our measurements suggest $\log_{10}M_{\rm cen}\simeq 12.00$, 12.05 and 12.15
($h^{-1}M_\odot$) for $\sigma=25$, 50 and 100 per cent,
corresponding to an average halo mass
$\langle \log_{10}\,M\rangle\simeq 11.9-12.0$ ($h^{-1}M_\odot$).

\begin{figure}
\begin{center}
\resizebox{3.4in}{!}{\includegraphics{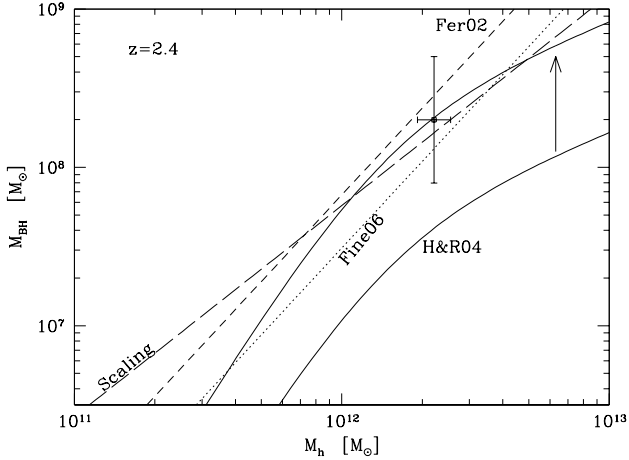}}
\end{center}
\caption{The halo-black hole mass relation.
The point with errors shows the mean and 1-$\sigma$ spread in $\log M_h$ for
our lognormal model (as in \S\protect\ref{sec:lognormal}) with 100 per cent
scatter.  The vertical error bars convert the luminosity range to $M_{BH}$
assuming $L=L_{\rm Edd}$.  The lines indicate $M_{BH}-M_h$ relations
inferred from the literature and from our scaling model.
The long dashed line gives the relation for the scaling model
(\S\protect\ref{sec:scaling}) which best fits our data.
The short dashed line is the result of \protect\citet{Fer02} and the dotted
line with the same slope is the result of \protect\citet{Fin06}.
The lower solid line indicates the local scaling relation of
\protect\citet{HarRix04} assuming $M_{\rm bulge}$ equals the stellar mass
inferred from the relation of \protect\citet{Mos10} at $z=2.4$.  The upper
solid line assumes $M_{BH}$ is 5 times larger at fixed $M_\star$ than the
local relation.  Note the curvature of the line due to the inefficiency of
galaxy formation at high and low halo mass.}
\label{fig:mbh_mhalo}
\end{figure}

The majority of high mass galaxies at high redshift are the central galaxy
in their dark matter halo, so observational stellar mass functions can
provide constraints on the stellar-to-halo-mass relation at $z\sim 2-3$
\citep[see][for recent examples]{Mos10,BehConWec10}.
In combination with the constraints on $M_{\rm cen}$ from quasar clustering,
and an assumption about the mean Eddington ratio of our sample, we can infer
the typical $M_{BH}-M_\star$ relation for our quasars.
Taking the lognormal model and adopting the conversion of \citet{Mos10},
the average stellar mass is $\log_{10}(M_\star/M_\odot)=10$--10.2
(with larger values corresponding to larger $\sigma$).
If the accretion occurs at $\lambda L_{\rm edd}$ the median black hole mass
is $\log_{10}(\lambda M_{BH}/M_\odot)=8.3$ using the conversions of
\citep{Cro05,Shen++09}.
We compare these numbers to a variety of published $M_{BH}-M_h$ relations
in Fig.~\ref{fig:mbh_mhalo}.
Our results are in broad agreement with the high redshift inferences but
predict larger black holes (at fixed halo mass) than what would be inferred
from the local relation of \citet{HarRix04}, even if we assume all of the
stellar mass associated with the halo central galaxy is in the bulge and
that quasars radiate at Eddington ($\lambda=1$).
This result argues that $M_{BH}$ should increase, at fixed $M_\star$, by a
factor of approximately $5\,\lambda^{-1}$ between $z=0$ and $z\simeq 2.4$.
This change is consistent with the increase measured in lensed quasar
hosts by \citet{Pen06} and the model of \citet{Hop07a}.
By comparison, the model of \citet[][Fig.~1]{Croton06} predicts roughly an
order of magnitude increase in $M_{BH}$ at $M_\star\sim 10^{10}$ between $z=0$
and $z=3$.  On the other hand, the simulations of \citet[][Fig.~15]{Sij07}
predict almost no evolution at the massive end.
\citet{Mer10} infer evolution of $M_{BH}-M_\star$ from the zCOSMOS survey
with a best-fit power-law of $(1+z)^{0.68}$
--a factor of 2.3 between $z=0$ and $2.4$--
while \citet{Dec10} measure a best-fit power-law of $(1+z)^{0.28}$
--a factor of 1.4 between $z=0$ and $2.4$-- from a carefully constructed sample
of 96 quasars drawn from the literature.
Our result favors stronger evolution, but given the statistical and systematic
uncertainties in all of the measurements, the uncertainties in stellar mass
estimates and selection biases towards more massive black holes in flux-limited
surveys all we can say is that it is encouraging that we see evolution in the
same {\em sense}.

Inverting this argument, we note that, at these high redshifts and
masses, obtaining a reasonable $M_{BH}-M_h$ relation provides constraints
on the possible occupancy distributions of quasars.  In particular, because
the halo mass function is much steeper than the galaxy stellar mass function,
the typical stellar mass of a central galaxy drops steeply with decreasing
halo mass---another way of stating that galaxy formation is inefficient in
low mass halos.
If black-hole properties are set by the galactic potential
rather than by halo properties we expect curvature in the $M_{BH}-M_h$
relation, which impacts how we interpret the duty cycle or the active quasar
fraction.

%[\citet{Tre02} has $M_{BH}=10^{8.13}M_\odot\ \sigma_{200}^4$.
%\citet{Fer02} has $v_c=10^{0.55}\sigma^{0.84}$.]

%\citet{BanCraSim09} do $M_{BH}-M_{\rm gal}$ where $M_{\rm gal}$ is
%$M_{200c}$ from SLACS.
%$\log_{10}(M_{BH})=8.18+1.55\left[\log_{10}(M_h)-13\right]$.
%Authors assume SIE out to $R_{200c}$.

%[\citet{DeGraf11} claim $\sigma\simeq 0.2$ in $\ln$ for a log-normal model
% of high-$z$ quasars.  Their fit corresponds to $L\sim M_h^{5/3}$ and
% they claim $z$-independence for $z\ge 3$.]

\begin{figure*}
\begin{center}
\resizebox{3in}{!}{\includegraphics{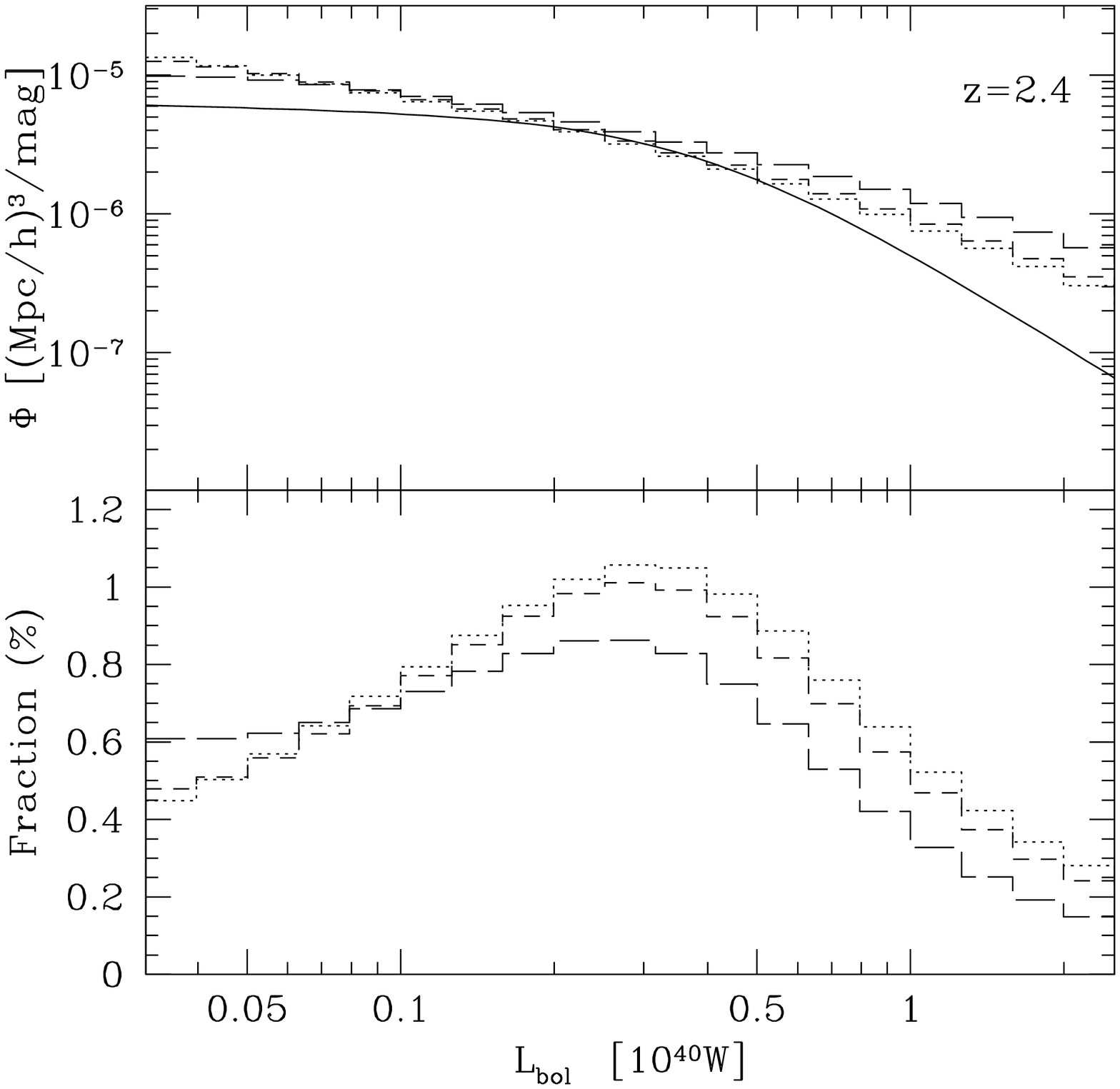}}
\resizebox{3in}{!}{\includegraphics{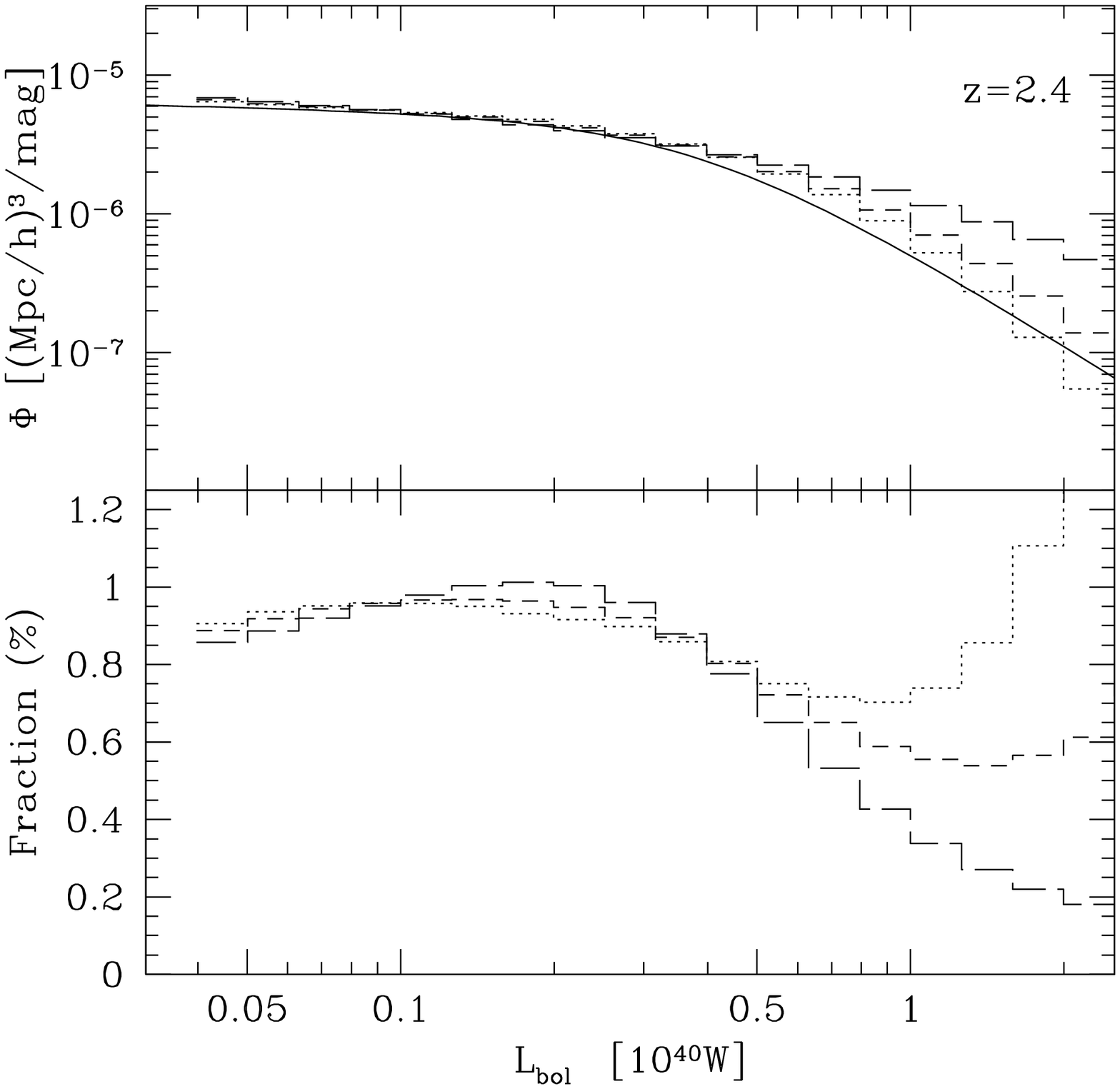}}
\end{center}
\caption{The duty cycle or active fraction of quasars.  The solid line is the
luminosity function of quasars from \protect\citet{Cro04}, as modified by
\protect\citet{Croton09}, converting from $b_J$ to $M_i$ as
$M_i(z=2)=M_{b_J}-0.71$ \protect\citep{Ric06} and from $M_i$ to $L_{\rm bol}$
via $M_i(z=2) = 72.5-2.5\log_{10}(L_{\rm bol})$ \protect\citep{Shen++09}.
The left hand plot shows, as the histograms in the upper panel, the predicted
luminosity functions (divided by 100) from the scaling relation model with
$L_{\rm bol}\sim v_{\rm peak}^4$ and 25, 50 and 100 per cent scatter,
assuming all black holes are active.
The lower panel shows the fraction that needs to be active at any given
time in order to obtain the observed luminosity function (solid line).
This is the duty cycle.
The right hand plot shows the same parameters for a model with
$L_{\rm bol}\sim M_{\rm gal}^{4/3}$ and in which we have randomly sampled
halos above $10^{12}\,h^{-1}M_\odot$ as described in the text.
The possible quasars in our simulations are downsampled by these duty cycles
to ensure a perfect match to the luminosity function as described in the text.}
\label{fig:duty}
\end{figure*}

\subsection{Scaling-relation model} \label{sec:scaling}

Another possibility is that the instantaneous luminosity of the quasar is
drawn from a lognormal distribution with central value proportional to a
power of the halo mass (or circular velocity) or the central galaxy mass
(or dispersion).  Physically such a model would arise if quasars radiate with
a small range of Eddington ratios and are powered by black holes whose masses
are tightly correlated with the mass or circular velocity of the galaxy or
host halo (through e.g.~black-hole bulge, bulge galaxy and galaxy halo
correlations).  The lognormal scatter is a combination of the dispersions in
each of the relations connecting instantaneous luminosity to halo mass
\citep[see e.g.][for recent examples of such models and Appendix
\ref{sec:halomodel} for further references]{Croton09,She09}.

We consider two examples here.  First we relate the black-hole properties to
those of the host halo directly.  We choose to use the peak circular velocity
of the dark matter halo as our measure of halo size and take $\log L$ to
be normally distributed around the (log of)
\begin{equation}
  L_{\rm pk}=L_0\left(\frac{v_{\rm peak}}{200\,{\rm km}\,{\rm s}^{-1}}\right)^4
  \quad .
\end{equation}
The normalization, $L_0$, is set by matching the clustering amplitude at a
given luminosity.
In principle we can allow the power-law index to vary\footnote{For example,
an index of $4$ would be appropriate to black holes whose growth is stopped
by momentum-driven winds and $5$ for those whose growth is stopped by
luminosity-driven winds \protect\citep{SilRee98}.}.
Unfortunately the range of luminosities we can probe observationally is
relatively small and the differences in index are difficult to measure.

\begin{figure}
\begin{center}
\resizebox{3.4in}{!}{\includegraphics{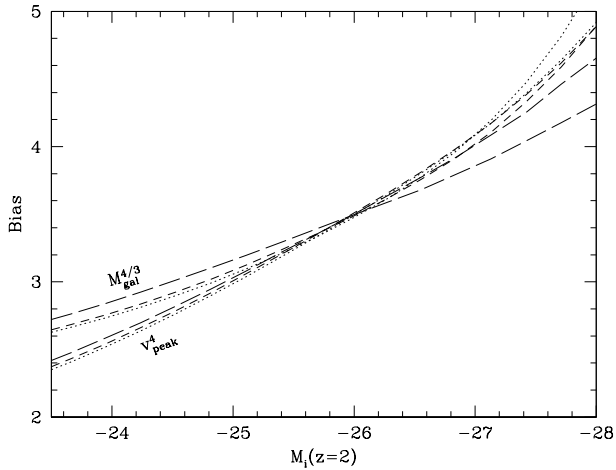}}
\end{center}
\caption{The predicted run of bias with magnitude for quasar samples two
magnitudes wide in $M_i(z=2)$ centered on the x-ordinate.  There are two
sets of curves, each with scatter in $L_{\rm obs}$ of $\sigma=25$ (dotted),
50 (dashed) and 100 per cent (long dashed).  The steeper curves
correspond to the model with $L_{\rm bol}\sim v_{\rm peak}^4$ while the
flatter curves correspond to $L_{\rm bol}\sim M_{\rm gal}^{4/3}$.  In
each case the active fraction has been adjusted to match the observed
luminosity function, as in Fig.~\protect\ref{fig:duty}.  The dependence
on the amount of scatter assumed is small because of the two magnitude
wide bin that is taken for each sample to increase statistics.}
\label{fig:bias_vs_Mi}
\end{figure}

For a given $L_0$ and scatter the halo population defines a luminosity
function of possible quasars.  The comparison of this to the observed
luminosity function of active quasars allows us to set the duty cycle,
which will be luminosity (and hence halo mass) dependent.
For each model we generate a mock catalog drawn from the halos of the
simulations introduced in \S\ref{sec:lognormal}.  We impose the duty cycle
by randomly subsampling the possible quasars to ensure the distribution
matches the observed luminosity function and then impose the magnitude
limits to match the observed sample.
We compute $w_p$ and fit for $r_0$ as described previously.

Fig.~\ref{fig:duty} shows the duty cycle for the best-fit model with
$\log_{10}L_0=38.8$.  The duty cycle peaks near one per cent at
$\log_{10}L_{\rm bol}\simeq 39.5$, corresponding to $M_i(z=2)\simeq -26.3$ or
black hole masses of $(2-3)\times 10^{8}M_\odot$.
This is near the center of our magnitude range and in our model corresponds
to halos of several times $10^{12}\,h^{-1}M_\odot$ or
$v_{\rm peak}\sim 300\,{\rm km}\,{\rm s}^{-1}$.
Converting the duty cycle into an activity time is somewhat ill defined.
If we assume $t_Q=f_{\rm on}t_H$, with $t_H$ the Hubble time, we
find $t_Q\sim 10^7\,$yr. 
These activity times are broadly consistent with those derived at $z \sim 0$,
though since the Hubble time is significantly shorter the duty cycles are
significantly higher. 
Also, note the luminosity/halo mass dependence of the duty cycle, which
implies that in this model we need extra physics to describe the LF
beyond simply major mergers with a fixed light-curve.

%\begin{figure}
%\begin{center}
%\resizebox{3in}{!}{\includegraphics{descend.eps}}
%\end{center}
%\caption{Halo occupation distribution for quasars and quasar descendants.
%The short-dashed line shows the occupation distribution for quasars in the
%scaling model at $z=2$.  The dotted line shows the distribution for the
%descendants of quasars above the same luminosity threshold at $z=3$.  The
%long-dashed line shows the distribution for quasars at $z=2$ above twice
%the luminosity threshold.}
%\label{fig:descend}
%\end{figure}

%\begin{table}
%\begin{center}
%\begin{tabular}{ll}
%${\rm log}M_{BH}=8.0+1.65 m$ & \protect\citep{Fer02} \\
%${\rm log}M_{BH}=7.8+1.82 m$ & \protect\citep{Fer02} \\
%${\rm log}M_{BH}=7.5+1.82 m$ & \protect\citep{Fin06}
%\end{tabular}
%\end{center}
%\caption{Halo mass-black hole mass relations.
%Here $m={\rm log}\,M_{180b}/(10^{12}M_\odot)$.  This table may go away, it is
%here for reference.}
%\label{tab:mbh_mhalo}
%\end{table}

A second option is to tie the black-hole properties to the host galaxy,
relating the galaxy properties to those of the halo by abundance matching.
For the stellar masses and dispersions of interest here the velocity
dispersion of the galaxy is proportional to the galaxy circular velocity,
and we can take the black-hole mass to scale as the $4^{\rm th}$ power of
either quantity \citep[as in the local universe;][]{Tre02}.  Then
\begin{equation}
  L_{\rm pk}=L_0\left(\frac{\sigma_\star}{200\,{\rm km}\,{\rm s}^{-1}}\right)^4
  \propto M_\star^p \quad , \quad p\approx 1
\end{equation}
where the power-law index is approximately unity in both observations and
numerical simulations
\citep[for representative examples see][and references therein]{HarRix04,Hop07a}.
Fig.~\ref{fig:duty} shows that in this model the low luminosity slope of the
luminosity function is in good agreement with the observations or the duty
cycle has little luminosity dependence---much of the suppression of low
luminosity quasars can be accomplished by the same physics as is invoked to
suppress star formation in lower mass galaxies.
This cut-off in the occupancy to low halo masses tends to flatten the run of
bias with luminosity (Fig.~\ref{fig:bias_vs_Mi}) in a manner similar to
luminosity dependent lifetime models, where there are also very few quasars
in low mass halos.

At the high luminosity end the suppression could be due to increasing
inefficiency in feeding a black hole as cold-mode accretion becomes less
effective \citep[e.g.][]{Sij07} or the fact that it is harder to have a
major galaxy merger (which would simultaneously drive gas to the center
and deepen the potential, allowing luminous quasar activity) when the
stellar mass is increasing slowly with halo mass or due to curvature in
the $M_{BH}-M_{\rm gal}$ relation \citep[e.g.][]{Gra12}.
To illustrate the general point we have randomly subsampled the halos above
$10^{12}\,h^{-1}M_\odot$ by a fraction $2\times 10^{12}/(M+10^{12})$
to obtain the duty cycles shown in the lower right panel of Fig.~\ref{fig:duty}.
While the agreement is by no means perfect, the general trends are in quite
good agreement with observations, i.e.~the luminosity dependence of the duty
cycle is relatively weak.

The determining factors for the flatness of the bias-luminosity relation
(Fig.~\ref{fig:bias_vs_Mi})
is the slope of the halo mass observable (i.e.~luminosity) relation and the
degree of scatter in that relation.
For very high clustering amplitudes (as measured at high $z$) one obtains an
upper limit on the scatter, which can be quite constraining for models, but at
intermediate $z$ quite a large degree of scatter is allowed
\citep{WhiMarCoh08,ShaWeiShe10}.
The scatter can arise from a number of sources including luminosity dependent
lifetime, but also stochasticity in the relations between halo mass and galaxy
mass, galaxy mass and central potential well depth, potential well depth and
black hole mass, black hole mass and optical luminosity.
For reasonable values of halo-observable slope, luminosity bin width and
stochasticity, one obtains quite flat $b(L)$ -- whether or not there is a
sharp cut-off in the halo mass distribution.
Thus while it is definitely plausible that quasar lifetime is luminosity
dependent, it is not strictly required by the current clustering data at
these redshifts.

\begin{figure}
\begin{center}
\resizebox{3.4in}{!}{\includegraphics{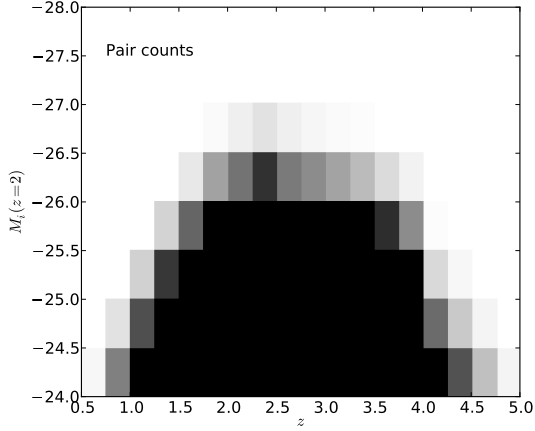}}
\end{center}
\caption{The number of quasar pairs with separation $<20\,h^{-1}$Mpc in the
redshift and magnitude bins shown, as predicted by the luminosity function
of \protect\citet{Cro04} as modified by \protect\citet{Croton09}.
We assume all quasars in the redshift and magnitude bin are observed over
$10^4$ sq.~deg.~of sky.  Reducing the sky area reduces the pair counts
linearly while finding only a fraction of the quasars reduces the counts
quadratically.
The black regions indicate more than $10^3$ pairs, the grey regions
indicate between $10^2$ and $10^3$ pairs and the white regions fewer than
$10^2$ pairs.
A solid detection of clustering from the auto-correlation should be possible in
the black regions while in the white regions it is likely necessary to
cross-correlate with more numerous samples or to make wider bins in redshift
or magnitude.}
\label{fig:pair_counts}
\end{figure}

The run of bias with halo mass becomes quite shallow at the low mass end.
If quasar luminosity is additionally affected by the inefficiency
of galaxy formation in lower mass halos, we expect to see a luminosity
dependent quasar bias primarily at higher luminosities (and redshifts).
Unfortunately this is where quasars become increasingly rare, which argues that
cross-correlation may be the best means of obtaining a strong clustering
signal \citep[e.g.][]{Shen++09}.
We explore this issue briefly in Fig.~\ref{fig:pair_counts}, where we show the
number of quasar pairs we expect to see in a complete, $10^4$ sq.~deg.~survey
with separation $<20\,h^{-1}$Mpc in bins of redshift and magnitude.
In the black areas, with more than $10^3$ pairs, a solid detection of
clustering from the auto-correlation should be possible.  In the other regions
(or for smaller or less complete surveys) cross-correlation with other quasars
or a different tracer is likely required.

By contrast the bias at the low luminosity end is an effective discriminator
between models where quasar luminosity depends on halo mass or galaxy mass.
Unfortunately BOSS is unable to probe this range of quasar luminosities.
Future surveys which probe fainter magnitudes may be able to settle this
question.  It may also be possible to cross-correlate faint, photometrically
selected quasars with the brighter spectroscopic quasars from the BOSS sample
\citep[using e.g., the methods in][]{MyeWhiBal09}.

\subsection{Redshift evolution} \label{sec:evolution}

\begin{figure}
\begin{center}
\resizebox{3.4in}{!}{\includegraphics{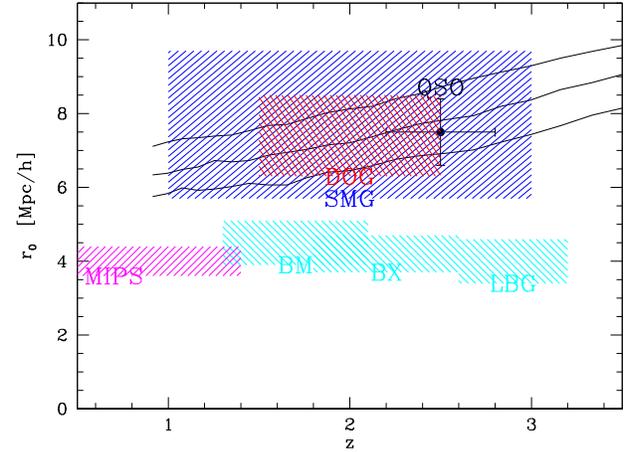}}
\end{center}
\caption{Comparison of our measured correlation length (square with error
bars) to that of other objects at $z\sim 2$.  In each case the horizontal
error bar or width of the shaded region shows the redshift range over which
the measurement is performed while the vertical error bar or height of the
shaded region shows the $\pm 1\,\sigma$ region.
The other data sets are mid-IR selected star-forming galaxies
\citep[MIPS;][]{Gil07}, BM, BX and LBG star-forming galaxies \citep{Ade05},
submillimetre galaxies \citep[SMG;][]{Hic12} and dust-obscured galaxies
\citep[DOG;][]{Bro08}.
The lines show the clustering of halos with peak circular velocity
$250\,$, $300$ and $350\,{\rm km}\,{\rm s}^{-1}$ as measured from the
simulations described in \S\protect\ref{sec:lognormal}.
See \protect\citet{Kru10}, Figs.~9-11, for a similar comparison at lower $z$.}
\label{fig:other_obj}
\end{figure}

In Fig.~\ref{fig:other_obj} we compared the clustering of our quasar sample
to that of other well-studied objects at $z\sim 2$.
The quasar clustering amplitude is similar to that of submillimetre galaxies,
suggesting they live in similar mass halos, and stronger than the typical star
forming population.
This is consistent with an evolutionary picture in which a merger triggers
a massive starburst that creates a submillimetre galaxy and which is then
quenched by the formation of a bright quasar \citep[e.g.][]{AleHic12}.
The descendants of our quasar hosts will have comparable clustering amplitudes
to the quasars themselves, indicating that they will likely evolve into
massive, luminous early-type galaxies at low redshift.

To better understand the possible fate of quasar host halos we employ another
high-resolution N-body simulation which allows us to track halos and subhalos
down to $z=0$ \citep[for details see][]{WhiCohSmi10}.
We select all halos at $z=2.4$ which are central subhalos of halos which
lie within an octave (i.e.~factor of 2) in mass centered on
$2\times 10^{12}\,h^{-1}M_\odot$ ($>10^4$ particles).
Essentially all of these halos are the most massive in their local environment.
One quarter of the hosts fall into a large halo (becoming a satellite) and
then lose more than 99.9 per cent of their mass (falling below the resolution
limit of the simulation) without merging with the central galaxy or another
satellite subhalo.  To the extent that this process is well resolved,
the stars in any galaxies hosted by these subhalos would likely contribute to
an intra-halo light component, while the black holes would form a freely
floating component or be associated with highly stripped satellites.
Of the remaining 75 per cent of the subhalos, three quarters are the most
massive progenitor in all their subsequent mergers and remain central galaxies
to $z=0$.  The remaining quarter become satellites which survive to $z=0$
inside more massive halos.  As a whole the population inhabits $z\simeq 0$
halos over a broad range of masses ($2\times 10^{12}-10^{15}\,h^{-1}M_\odot$)
peaking at $(1-2)\times 10^{13}\,h^{-1}M_\odot$.
Halos of this group scale host galaxies of a few $L_\star$ today, a population
dominated by elliptical galaxies.
This wide diversity of outcomes is reminiscent of the varied fates of
$z\sim 2$ star-forming galaxies \citep[e.g.][]{Con08}.
As constraints on the stellar masses of galaxies within halos become tighter,
comparison of the stellar masses of quasar hosts with that of their $z\sim 0$
descendants will put constraints on the star-formation history of these
objects.

The evolution of clustering with time can place strong constraints on how
episodic quasar activity can be.  As emphasized by \citet{Cro05}, if the
typical host of quasars does not evolve significantly with redshift then
quasars cannot be repeated bursts of the same black hole because such black
holes would live in halos which grow in mass as the Universe evolves.
However the quantitative strength of this statement, and the allowed fraction
of objects which could burst more than once, is difficult to assess.
One issue is the size of the observational errors, the other is that higher
redshift samples typically probe more massive black holes than lower redshift
samples \citep[as emphasized e.g.~by][]{Hop07b}.

For example, the highest redshift clustering measurement comes
from \citet{She07}, at $z\simeq 4$.  Their result is
consistent with host halos of $(2-10)\times 10^{12}\,h^{-1}M_\odot$.
Such hosts would grow in mass by a factor of approximately 4
between $z=4$ and 2.4.  Similarly, the progenitors of our
$2\times 10^{12}\,h^{-1}M_\odot$, $z=2.4$ halos are approximately
4 times less massive at $z=4$.  This suggests on average an order
of magnitude mismatch in halo masses, but with a large error.
Assuming no evolution in Eddington ratio between $z=4$ and
$z=2.4$ the quasars in the \citet{She07} sample are a factor
of about 5 brighter than the BOSS quasars.  If the power-law
index of the $M_{BH}-M_h$ relation is close to unity there is
little tension from clustering\footnote{The dramatic decrease in
quasar numbers to higher redshift does impose constraints.} in assuming
quasars episodically burst.

The time span between $z=4$ and $z=2.4$ in our adopted cosmology
is $1\,h^{-1}$Gyr.  The Universe is another $1\,h^{-1}$Gyr older
at $z\simeq 1.5$.  Taking $r_0$ at $z=1.57$ from \citet{Ros09}
and converting it to a host halo mass we obtain
$\sim 5\times 10^{12}\,h^{-1}M_\odot$.
The descendants of our $2\times 10^{12}\,h^{-1}M_\odot$, $z=2.4$ halos
are approximately twice as massive at $z=1.5$, again leading to
little tension in a model with episodic outbursts.

We can see the issues most clearly in Fig.~\ref{fig:prevwork}.
If halos moved with the same large-scale velocities as the dark matter
and never merged the decrease of their bias would approximately cancel
the increased clustering of the matter (\citealt{Fry96}; see Fig.~1 of
\citealt{NDWFS}).
For a highly biased sample such passive evolution corresponds to almost
constant clustering strength, for slightly less-biased objects the clustering
strength grows slowly with time.
This result is shown as the dashed line in Fig.~\ref{fig:prevwork},
corresponding to objects with $b(z=0)=1.8$.
Merging of halos and including a finite range of halo masses alters the
details of this evolution, but keeps the sense unaltered.
By contrast, halos of a fixed mass predict a clustering strength which drops
slowly with time -- shown as the dotted line in Fig.~\ref{fig:prevwork} for
halos of $2\times 10^{12}\,h^{-1}M_\odot$.
Again, a more realistic scenario with a finite range of halo masses has the
same sense of the evolution.
The current data are not in strong conflict with either scenario.
If a random fraction of quasars repeats at later epochs while new quasars
always appear in halos of a fixed mass, the measured clustering resembles a
sample with bias $f_{\rm repeat}b_{\rm repeat}+(1-f_{\rm repeat})b_{\rm new}$.
The measured values of $r_0$ are consistent with being roughly constant over
$1<z<3$.
For this reason it is difficult to put a strong upper limit on the fraction
of quasars that turn on more than once or the maximum number of times a
given quasar can burst.

%\citet{Lid06} argue that non-evolution of the characteristic mass comes from
%a cancellation of two trends:
%\begin{itemize}
%\item Characteristic $M_h$ comes from turnover in $L_{\rm peak}$ distribution
%and $L_{\rm peak}(M_h)$ relation.  The turnover in $L_{\rm peak}$ is basically
%that $f_{\rm on}$ looks lognormal-ish in order to get $\Phi(L)$ right.
%\item Relation of $M_h$ to $L_{\rm peak}$ evolves through $M_{BH}-M_h$.
%\item Turnover in $L_{\rm peak}$ distribution comes from break in observed QLF.
%\end{itemize}

%\subsection{Threshold crossing model} \label{sec:threshold}
%Objects which cross a fixed mass threshold plus fuel exhaustion.

\section{Conclusions}
\label{sec:conclusions}

We have performed real- and redshift-space clustering measurements of
a uniform subsample of quasars observed by the Baryon Oscillation Spectroscopic
Survey (BOSS).  These quasars lie in the redshift range $2.2<z<2.8$
where there has previously been a gap in coverage due to the fact that the
quasar and stellar loci cross and the difficulty in targeting quasars in
large numbers to faint fluxes.
We detect clustering at high significance in both real- and redshift-space
for the entire sample and subsamples split by redshift and luminosity
(see Table \ref{tab:results}).
We do not detect a luminosity or redshift dependence of the clustering
strength, although our sensitivity to this dependence is weak due to the
limited dynamic range in both variables in our sample.

The two-point correlation functions are consistent with power-laws over the
range of scales measured, with an underlying real-space clustering of the
form $\xi(r)=(r_0/r)^2$.
The correlation length, $r_0$, does not appear to evolve strongly over the
redshift range $z\simeq 3$ to $1$.  This result is consistent with passive
evolution of a highly biased population, although this interpretation is by
no means unique.
Our results are consistent with quasars living in halos of typical mass
$10^{12}\,h^{-1}M_\odot$ at $z\simeq 2.4$, in line with expectations from
earlier surveys.
The measured bias and space density of quasars can be used to infer their
duty cycle \citep{ColKai89}.  For our best-fit models the duty cycle peaks
at one per cent, implying an activity time of $\sim 10^7$ years.
This time is comparable to the activity times inferred for quasars at lower
redshift, although the Hubble time at $z=2.4$ is shorter than at lower
$z$ and hence the active fraction is larger.

The typical host halo mass is similar to the inferred hosts of submillimetre
galaxies and is more massive than the inferred hosts of typical star-forming
galaxies at the same redshift.  This interpretation is in turn consistent
with an evolutionary picture in which a massive starburst creates a
submillimetre galaxy and is quenched by the formation of a bright quasar.
While the typical descendant of the halos that host BOSS quasars is likely to
host a luminous, elliptical galaxy at the present time, we find a wide
diversity in descendants in N-body simulations.

Using abundance matching to infer the properties of quasar host galaxies
we find evidence for evolution in the $M_{BH}-M_{\rm gal}$ relation in
the sense that black holes must be $\approx 5\times$ more massive at fixed
galaxy mass at $z=2.4$ than at $z\simeq 0$.
We find that the predictions for how quasar activity and clustering (bias)
depend on luminosity differ depending on whether we take as our fundamental
relationship a black hole-halo correlation or a black hole-galaxy correlation.
This is because the efficiency of galaxy formation is strongly (halo) mass
dependent for the halos of interest at these redshifts, leading to strong
curvature in the $M_{\rm gal}-M_h$ relation.
In either scenario a modest scatter between halo or galaxy mass and
observed quasar luminosity (arising, for example, from a combination of
scatters in the black-hole bulge, bulge galaxy and galaxy halo correlations
and the Eddington ratio) leads to a shallow dependence of clustering on
quasar luminosity, as observed.

Future surveys of quasars which probe different regions of the
luminosity--redshift plane will inform models of quasar formation
and evolution.  In this regard BOSS continues to measure quasar
redshifts, and we expect the number of quasars in the luminosity and
redshift range discussed here will be more than doubled by the end
of the survey.
It may be possible to incorporate the additional, BONUS quasars
through cross-correlation or to cross-correlate spectroscopic and photometric
quasar samples to better allow us to break the sample by luminosity,
spectral or radio properties.
In addition BOSS is measuring redshifts for a large sample of $z>3$ quasars,
and analysis of those data will be crucial in understanding the early phases
of quasar growth.

\medskip

  Funding for SDSS-III has been provided by the Alfred P. Sloan Foundation,
  the Participating Institutions, the National Science Foundation, and the
  U.S. Department of Energy Office of Science.
  The SDSS-III web site is http://www.sdss3.org/.

  SDSS-III is managed by the Astrophysical Research Consortium for the
  Participating Institutions of the SDSS-III Collaboration including the
  University of Arizona,
  the Brazilian Participation Group,
  Brookhaven National Laboratory,
  University of Cambridge,
  Carnegie Mellon University,
  University of Florida,
  the French Participation Group,
  the German Participation Group,
  Harvard University,
  the Instituto de Astrofisica de Canarias,
  the Michigan State/Notre Dame/JINA Participation Group,
  Johns Hopkins University,
  Lawrence Berkeley National Laboratory,
  Max Planck Institute for Astrophysics,
  Max Planck Institute for Extraterrestrial Physics,
  New Mexico State University,
  New York University,
  Ohio State University,
  Pennsylvania State University,
  University of Portsmouth,
  Princeton University,
  the Spanish Participation Group,
  University of Tokyo,
  University of Utah,
  Vanderbilt University,
  University of Virginia,
  University of Washington,
  and Yale University.

  The analysis made use of the computing resources of the
  National Energy Research Scientific Computing Center,
  the Shared Research Computing Services Pilot of the
  University of California and
  the Laboratory Research Computing project at
  Lawrence Berkeley National Laboratory.

  M.W. was supported by the NSF and NASA.
  A.D.M. is a research fellow of the Alexander von Humboldt Foundation
  of Germany.
  JM is supported by Spanish grants AYA2009-09745 and PR2011-0431.
  JB was partially supported by a NASA Hubble Fellowship (HST-HF-51285.01)

\appendix

\section{Redshift errors}
\label{sec:zerror}

At the signal-to-noise ratio and redshift at which BOSS is working it is
difficult to obtain precise redshifts for quasars.  Emission lines, such as
Mg{\sc ii} which are good redshift indicators and which can be used at lower
redshift, have redshifted into a relatively noisy part of the spectrum or off
of the device altogether.
One hour integrations on a $2.5\,$m telescope make it difficult to
measure redshifts for quasars with weak lines.

The BOSS pipeline measures quasar redshifts by fitting their spectra to a set
of PCA templates \citep{DR8} plus a cubic polynomial to allow for changes in
continuum slope.
The reduced $\chi^2$ vs.~redshift is mapped in steps of
$\Delta\log_{10}(\lambda)=10^{-4}$ from
$z=0.0033$ to $7$ and the template fit with the best reduced $\chi^2$ is
selected as the redshift.  In addition redshifts are computed by fitting any
lines in two groups (forbidden and allowed).
A comparison of redshifts determined from different lines \citep{Hen06b,She07},
and visual inspections, suggests the automatic redshifts are good to
$\Delta z/(1+z)\simeq 0.003$.
At $z=2.5$ this corresponds to an error in the line-of-sight distance of
roughly $10\,h^{-1}$Mpc (comoving), which is significant compared to the
correlation length of quasar clustering.
We are attempting to improve our quasar redshift determination, but for now
we simply account for the residual line-of-sight smearing induced by redshift
errors in our fitting.

\begin{figure}
\begin{center}
\resizebox{3.4in}{!}{\includegraphics{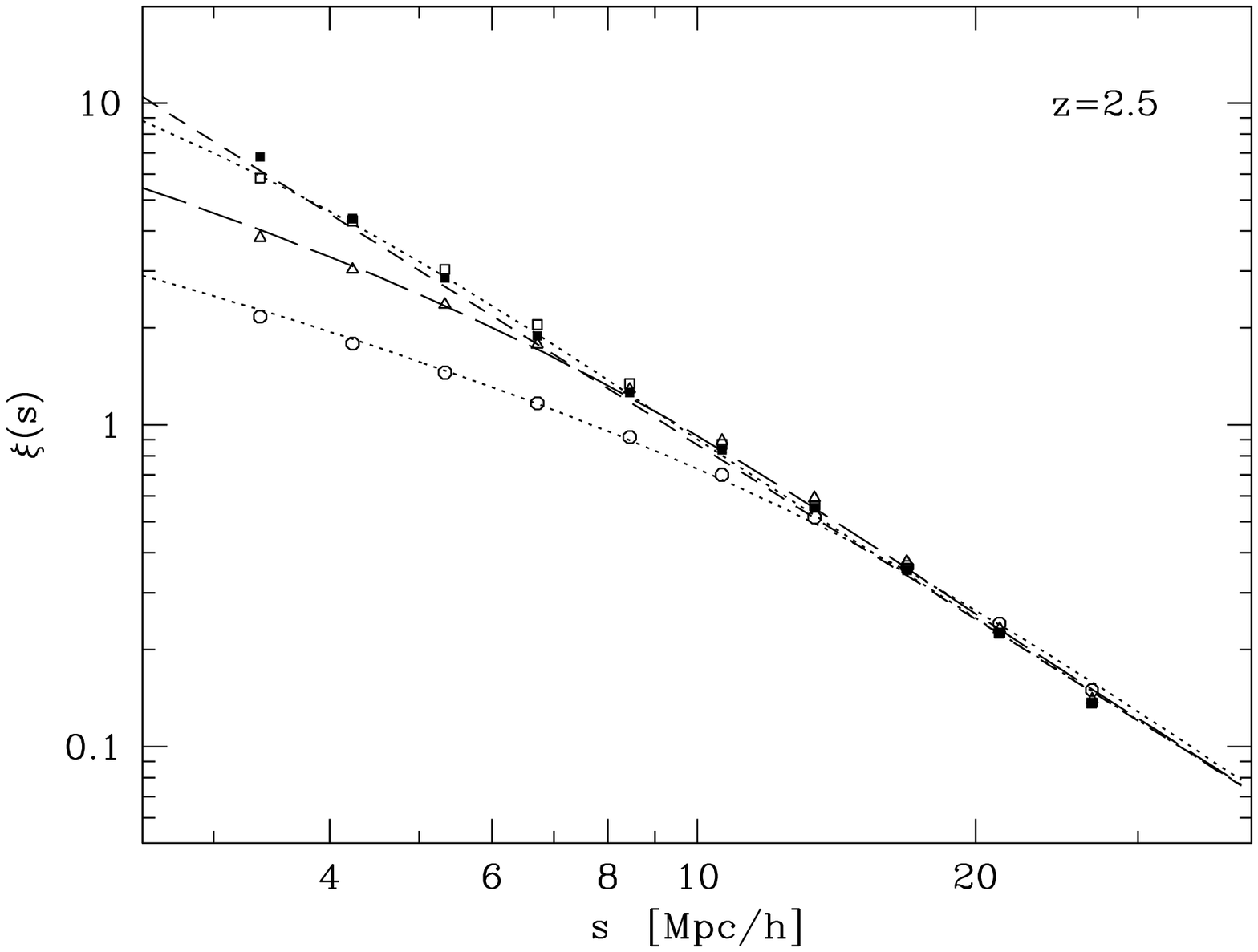}}
\end{center}
\caption{The effects of redshift errors on the redshift space (monopole)
correlation function.  The dashed line shows a power-law correlation
function with slope 1.8, and the other lines are for Eq.~(\ref{eqn:xi_smear})
with per-object redshift errors corresponding to 2.5, 5 and $10\,h^{-1}$Mpc.
The points are for mock quasars as in \S\protect\ref{sec:scaling}
with Gaussian errors added to the line-of-sight velocities.}
\label{fig:zerror}
\end{figure}

In the limit $b\gg 1$ the redshift-space halo correlation function is
approximately isotropic and a power-law.  If the redshift errors
on quasars are uncorrelated and Gaussian distributed with
fixed amplitude $\sigma_z$ the observed correlation function is
\begin{equation}
  \xi_{\rm obs}(s) = \frac{1}{2}\int \frac{dz}{s}
  \int\frac{dZ}{\sqrt{2\pi}\sigma}\ \xi\left(\sqrt{s^2-z^2},Z\right)
  e^{-(z-Z)^2/2\sigma^2}
\label{eqn:xi_smear}
\end{equation}
where $\sigma=\sqrt{2}(c\sigma_z)/H(z)$.  The integral, $\xi_{\rm obs}$,
divided by the input power-law is what we refer to in the main text as
$F(s)$.

Fig.~\ref{fig:zerror} compares this model, with a power-law correlation
function of slope 1.8, to halos from N-body simulations with applied Gaussian
line-of-sight velocity errors.  As expected, the agreement is excellent.
In this test the redshift errors were all drawn from a Gaussian of the same
dispersion.  In the observational sample we might reasonably expect the errors
to depend on the properties of each quasar.  In this situation we should
interpret $\sigma$ as a pair-weighted, ``effective'', redshift error.
The above tests, plus the clustering measurements themselves, are consistent
with a per quasar redshift error corresponding to $10\,h^{-1}$Mpc, and we use
that as our fiducial value throughout.

\section{Quasar model}
\label{sec:halomodel}

Models of the quasar phenomenon come in several basic flavors, but are in
fact all quite similar.  The majority of models assume that quasar activity
occurs due to the major merger of two gas-rich galaxies, since this scenario
provides the rapid and violent event needed to funnel fuel to the center of
the galaxy \citep[e.g.~via the bars-within-bars instability][]{Shl89} and
feed the central engine while at the same time giving a
connection between black hole fueling and the growth of a spheroidal stellar
component.
If black hole growth is feedback limited, it is only a rapidly growing
potential well that can host a rapidly accreting hole.
A notable exception is the models of \citet{Cio97,Cio01} which postulate the
fuel is funneled to the center by thermal instabilities, which provide the
rapid growth of the spheroidal component necessary for black hole growth.
As we shall see, both sets of models can predict very similar halo occupancy.

Some models implement the physics directly in numerical simulations which
attempt to track the hydrodynamics of the gas, with subgrid models for quasar
and star formation and the associated feedback
\citep{Sij07,Hop08,DeGraf11}.
Other models work at the level of dark matter halos, but follow the same
physics -- in simplified form -- semi-analytically
\citep{CatHaeRee99,KauHae00,KauHae02,VolHaaMad03,BroSomFab04,Granato04,
Croton06,MonFonTaf07,Mal07,Bon09,Fan12}.
Even more idealized are models which are built upon dark matter halos but
use scaling relations or convenient functional forms to relate the quasar
properties to those of their host halos
\citep{EfsRee88,Car90,WyiLoe02,WyiLoe03,HaiCioOst04,Mar06,Lid06,
Croton09,She09,BooSch10}.
A final level of abstraction is to simply provide a stochastic recipe for
populating dark matter halos with quasars which is tuned to reproduce the
observations as best as possible while not attempting to follow the underlying
physics
\citep{PorMagNor04,WhiMarCoh08,PWNP,ShaWeiShe10,VolSta11,Kru12,Kir12}.

The modeling is simplified by several facts.  Quasars are rare, their
activity times are short and the fraction of binary quasars is small
\citep{Hen06a,Mye07b}.
The hydrodynamic and semi-analytic models tend to reproduce the observed,
$z=0$ relation between black hole mass and halo, which is used as input to
the scaling models.  The models agree on the level of scatter in the
relation (roughly a factor of two) and in broad brush on the evolution in the
amplitude and slope of this relation with redshift.  Some models invoke
quasar feedback limited accretion explicitly while others achieve the same
scaling relations without such a limit---e.g., by coupling the mechanisms
by which bulges and black holes grow.
At these masses and redshifts the satellite galaxy fraction is tiny,
so the halo to stellar mass relation is set by abundance matching and any
model which reproduces the galaxy stellar mass function will reproduce this
relation.
At high redshift bright quasars radiate near Eddington, so the models predict
similar halo occupancies.

The probability that a halo will undergo a major merger in a short redshift
interval is only weakly dependent on the mass of the halo
\citep{LacCol93,Per03,CohWhi05,WetCohWhi09,FakMa09,Hop10}, i.e.~the mass
function of such halos is almost proportional to the mass function of the
parent population.
Similarly the clustering properties of recently merged halos are
similar to a random sample of the population with the same mass
distribution \citep{Per03,WetCohWhi09}.
Thus in any interval, $\Delta z$, the fraction of halos of mass $M_h$
which undergo a quasar event is almost independent of $M_h$ and
$z$ and can be regarded as a random selection.
This makes it difficult to infer that quasars arise from mergers simply from
their large-scale clustering, but also implies that for the purposes of
modeling the 1- and 2-point functions of the quasar population it is
sufficient to specify the halo occupation of the parent population
(more complex models may be needed if correlations between quasars
 and properties of e.g.~galaxies were required).

To determine whether a quasar candidate makes it into any sample it is
necessary to relate the observed luminosity to the peak luminosity
which is determined by $M_{BH}$.  This is done either by specifying a
light curve (e.g.~a power-law, a power-law with a varying slope or an
exponential) or directly $P(L|M_h)$ using the fact that the
triggering rate is understood.  For constant triggering rate a light curve
$L\sim t^{-\gamma}$ implies $P(L)\sim L^{-(\gamma+1)/\gamma}$, so a wide
range of $\gamma$ maps to a narrow range in $P(L)$ index.
For the high luminosity thresholds we only see objects very near their peak
brightness.  Taking into account the factor $\sim 2$ scatter in $M_{BH}$ at
fixed bulge mass, we expect $P(L|M_h)$ to be roughly lognormal with a similar
width (it can be slightly broader due to variation in the Eddington ratio at
peak).
As we probe lower luminosities we are more likely to see older, more
massive black holes leading to a low $L$ tail in $P(L|M_h)$ and more
lower-Eddington-ratio objects \citep{Lid06,She09,Cao10}.
Unless we cut off the probability that a halo hosts a quasar at low and
high halo mass, we will overproduce low and high $L$ sources
\citep{Lid06,Croton09,She09}.
In the physical models these limits occur due to lack of fuel in low mass
halos and the inability of gas to cool in high-mass halos
\citep[e.g.][]{CatHaeRee99}.
As discussed in \S\ref{sec:scaling}, models which match the $M_h-M_\star$
relation for galaxies tend to roughly match the required quasar suppression.

We are interested in the probability that a given halo hosts a quasar in
our sample, e.g.~$f_{\rm on}(M_h)\propto P(>L_{\rm min}|M_h)$.
{}From the arguments above we expect this relation to be an approximately
lognormal function which is asymmetric towards high $M_h$ at low $L$.
It is quite difficult to put constraints on the detailed form of this function
using luminosity function and clustering measurements.

We argued above that the steepness of the halo mass function and the high
bias of quasars implies that the quasar satellite fraction is small.  The
luminosity function thus provides a constraint on the duty cycle.  (This
extra degree of freedom reduces the ability of large-scale structure
measurements to constrain the halo occupancy, compared to modeling galaxies.)
The steeply falling mass function also implies that the number of quasars
hosted in very massive halos is small regardless of the occupancy statistics
of such halos.
On scales larger than the virial radius of the typical quasar host halo
(i.e.~$200-300\,h^{-1}$kpc) the 2-point function is dominated by pairs of
quasars in different halos, and thus primarily measures the quasar-weighted
halo bias which allows us to infer the mean, quasar-weighted halo mass.
This result remains true for cross-correlation studies too, provided we work
on scales larger than the virial radius of the quasar hosts.
On smaller scales the amplitude and slope of the correlation function allow
us to measure a combination of the satellite fraction of quasars and the low
mass cutoff.

\end{document}